\documentstyle[psfig,times]{mn}

\newif\ifAMStwofonts

\def\etal{et al.\ }

\def\ie{{\sl i.e.\ }}
\def\c{{\sl Chandra}}

\def\approxlt{\mathrel{\spose{\lower 3pt\hbox{$\sim$}}
        \raise 2.0pt\hbox{$<$}}}
\def\approxgt{\mathrel{\spose{\lower 3pt\hbox{$\sim$}}
        \raise 2.0pt\hbox{$>$}}}
        
\def\cm{{\rm\thinspace cm}}

\def\deg{$^\circ$}
\def\erg{{\rm\thinspace erg}}

\def\ha{H$\alpha$}
\def\hb{H$\beta$}

\def\keV{{\rm\thinspace keV}}

\def\km{{\rm\thinspace km}}
\def\kpc{{\rm\thinspace kpc}}

\def\Mpc{{\rm\thinspace Mpc}}
\def\Msun{\hbox{$\rm\thinspace M_{\odot}$}}

\def\s{{\rm\thinspace s}}
\def\yr{{\rm\thinspace yr}}

\def\ergpspcmsqpA{\hbox{$\erg\s^{-1}\cm^{-2}$\AA$^{-1}\,$}}

\def\ergps{\hbox{$\erg\s^{-1}\,$}}

\def\kmps{\hbox{$\km\s^{-1}\,$}}

\def\Msunpyr{\hbox{$\Msun\yr^{-1}\,$}}

\def\pcmsq{\hbox{$\cm^{-2}\,$}}

\def\kmpspMpc{\hbox{$\kmps\Mpc^{-1}$}}

\ifoldfss
  \ifCUPmtlplainloaded \else
    \NewTextAlphabet{textbfit} {cmbxti10} {}
    \NewTextAlphabet{textbfss} {cmssbx10} {}
    \NewMathAlphabet{mathbfit} {cmbxti10} {} 
    \NewMathAlphabet{mathbfss} {cmssbx10} {} 
  \fi
  \ifAMStwofonts
    \ifCUPmtlplainloaded \else
      \NewSymbolFont{upmath} {eurm10}
      \NewSymbolFont{AMSa} {msam10}
      \NewMathSymbol{\upi}     {0}{upmath}{19}
      \NewMathSymbol{\umu}     {0}{upmath}{16}
      \NewMathSymbol{\upartial}{0}{upmath}{40}
      \NewMathSymbol{\leqslant}{3}{AMSa}{36}
      \NewMathSymbol{\geqslant}{3}{AMSa}{3E}

      \let\leq=\leqslant \let\le=\leqslant
       
    \fi
  \fi
\fi 

\ifnfssone
  \newmathalphabet{\mathit}
  \addtoversion{normal}{\mathit}{cmr}{m}{it}
  \addtoversion{bold}{\mathit}{cmr}{bx}{it}
  \newmathalphabet{\mathbfit} 
  \addtoversion{normal}{\mathbfit}{cmr}{bx}{it}
  \addtoversion{bold}{\mathbfit}{cmr}{bx}{it}
  \newmathalphabet{\mathbfss} 
  \addtoversion{normal}{\mathbfss}{cmss}{bx}{n}
  \addtoversion{bold}{\mathbfss}{cmss}{bx}{n}
  \ifAMStwofonts
    \ifCUPmtlplainloaded \else
      \UseAMStwoboldmath
      \makeatletter
      \new@mathgroup\upmath@group
      \define@mathgroup\mv@normal\upmath@group{eur}{m}{n}
      \define@mathgroup\mv@bold\upmath@group{eur}{b}{n}
      \edef\UPM{\hexnumber\upmath@group}
      \new@mathgroup\amsa@group
      \define@mathgroup\mv@normal\amsa@group{msa}{m}{n}
      \define@mathgroup\mv@bold\amsa@group{msa}{m}{n}
      \edef\AMSa{\hexnumber\amsa@group}
      \makeatother
      \mathchardef\upi="0\UPM19
      \mathchardef\umu="0\UPM16
      \mathchardef\upartial="0\UPM40
      \mathchardef\leqslant="3\AMSa36
      \mathchardef\geqslant="3\AMSa3E

      \let\leq=\leqslant \let\le=\leqslant

    \fi
  \fi
\fi 

\ifnfsstwo
  \DeclareMathAlphabet{\mathbfit}{OT1}{cmr}{bx}{it}
  \SetMathAlphabet\mathbfit{bold}{OT1}{cmr}{bx}{it}
  \DeclareMathAlphabet{\mathbfss}{OT1}{cmss}{bx}{n}
  \SetMathAlphabet\mathbfss{bold}{OT1}{cmss}{bx}{n}
  \ifAMStwofonts
    \ifCUPmtlplainloaded \else
      \DeclareSymbolFont{UPM}{U}{eur}{m}{n}
      \SetSymbolFont{UPM}{bold}{U}{eur}{b}{n}
      \DeclareSymbolFont{AMSa}{U}{msa}{m}{n}
      \DeclareMathSymbol{\upi}{0}{UPM}{"19}
      \DeclareMathSymbol{\umu}{0}{UPM}{"16}
      \DeclareMathSymbol{\upartial}{0}{UPM}{"40}
      \DeclareMathSymbol{\leqslant}{3}{AMSa}{"36}
      \DeclareMathSymbol{\geqslant}{3}{AMSa}{"3E}

      \let\leq=\leqslant \let\le=\leqslant

    \fi
  \fi
\fi 

\ifCUPmtlplainloaded \else
  \ifAMStwofonts \else 
    \def\upi{\pi}
    \def\umu{\mu}
    \def\upartial{\partial}
  \fi
\fi

\title{The giant H$\alpha$/X-ray filament in the cluster of galaxies A\,1795}
\author[C. S. Crawford et al  ]
       {C. S. Crawford,\thanks{E-mail: csc@ast.cam.ac.uk} J. S. Sanders and A. C. Fabian\\
       Institute of Astronomy, Madingley Road, Cambridge CB3 0HA}

\date{Submitted to MNRAS January 2004}

\pubyear{2002}

\begin{document}

\maketitle

\label{firstpage}

\begin{abstract} The cluster of galaxies A\,1795 hosts a 46\kpc-long
filament at its core, which is clearly visible in the light of \ha\
and X-ray emission.  We present optical slit spectroscopy and deeper
{\sl Chandra} X-ray data of the filament. The optical spectra reveal
that the the bulk of the filament is quiescent (with
$\sigma<$130\kmps), although considerable velocity structure is 
apparent around the powerful radio source in the central cluster
galaxy, where a direct interaction is occurring between the radio
plasma and the surrounding intracluster medium.  The filament contains
a clump of UV/blue continuum halfway along its length, which we
resolve into a chain of at least 5 distinct knots using archival HST
images; the optical spectrum of this clump confirm it to be mostly
comprised of O stars.  It is well-removed from the central galaxy and
radio source, and is most likely an example of a group of young star
clusters condensing directly from the cooling gas in the filament.
The observed spatial offset between these knots of star formation and
the peak in the optical line emission confirms that the massive star
formation is most unlikely to be responsible for the bulk of the
observed emission-line luminosity in the filament. Some other (as yet
undetermined) source of energy is required to power and maintain the
optical line-emission, yet it must not completely impede the cooling
of the X-ray gas within the filament to form the star clusters.
\end{abstract}


\section{Introduction}

A\,1795 is a well-studied rich cluster of galaxies lying at a redshift
of $z=0.063$. The dominant central galaxy is of particular interest,
as it has long been known to possess a luminous, extended
emission-line nebula (Cowie \etal 1983; van Breugel, Heckman \& Miley
1984; Hu, Cowie \& Wang 1985; Heckman \etal 1989) as well as localized
regions of excess blue continuum indicative of massive recent star
formation (McNamara \& O'Connell 1992, 1993; Johnstone, Fabian \& Nulsen
1987; Cardiel, Gorgas \& Salamanca 1998; McNamara \etal 1996; Pinkney
\etal 1996). The nebula is in two major parts: one is about 10 kpc in
size, located within the central galaxy, and around its radio source;
the other is in the form of a giant filament extending beyond the
galaxy. 
The central galaxy hosts the powerful FR~I radio source
4C~26.42 (1346+268), which shows high Faraday rotation measures (Ge \&
Owen 1993). The radio emission is confined to within $10$~arcsec of the
active nucleus, and is distorted into a Z-shaped morphology with both
lobes bent through 90 degrees within $\pm$2~arcsec of the nucleus. A
bright hotspot is located less than one arcsecond to the NW of the
nucleus, before the bent \lq knee' of the northern radio lobe (Ge \&
Owen 1993; this is apparent only as a slight extension to the radio
nucleus in the radio morphology contours shown as part of
Fig~\ref{fig:zoomimages}).

The spatial distribution of the excess blue continuum and the optical
emission-line gas close to the galaxy is clearly related to that of
the radio source and its depolarization, suggesting a close physical
relationship. The brightest off-nuclear knot of line emission lies
along the outer boundary of the radio lobe to the north-west (van
Breugel et al 1984, Cowie et al 1983; Fig~\ref{fig:zoomimages}) and
the two primary regions of excess blue continuum light each curve
along the outside of a radio lobe (McNamara et al 1996). This UV/blue
light is resolved into a knotty structure by the HST, strongly
suggesting it to be due to clusters of young stars (McNamara et al
1996; Pinkney et al 1996; O'Dea et al 2004; 
Fig~\ref{fig:zoomimages}).  This
interpretation is supported by UV imaging used to constrain star
formation models, implying either continuous star formation or a
recent starburst, with rates of around $\sim5-20$\Msunpyr (Smith et al
1997; Mittaz et al 2001). The general inference from all these papers
is that close to the galaxy, recent star formation has been triggered
by the interaction between the radio jet with dense gas to the
north-west of the galaxy (eg McNamara et al 1996; Pinkney et al 1996).
A detection of CO emission from cold molecular gas in the central
galaxy (Edge 2001) lends further strength to this explanation, as a
large fraction of the CO emission is also located along this
north-west edge of the northern radio lobe (Salom\'e \& Combes 2004 :
Fig~\ref{fig:zoomimages}).  Colour maps also reveal a dust lane
sweeping down from the western edge of the northern radio lobe to
cross the galaxy nucleus (McNamara \etal 1996; Pinkney \etal 1996;
Fig~\ref{fig:zoomimages}).  Thus local to the radio source there is a
clear interplay between the outflowing jets and the cooler components
of the surrounding intracluster medium -- collision, displacement and
jet-induced star formation. 

The presence of features such as young massive star formation, an
extended luminous emission-line nebula and central cold molecular gas
are characteristic of central galaxies in clusters with a strong
cooling flow operating in the hot intracluster medium (Crawford 2003,
1999; Fabian 1994 and references therein). The most recent X-ray data
do indicate a cooling flow would operate in the absence of any heating
source within the inner $\sim200$\kpc\ of the A\,1795 cluster, with an
integrated mass deposition rate of $\sim100$\Msunpyr (Ettori \etal
2002; Tamura et al 2001; Molendi \& Pizzolato 2001).  Although the
X-ray emission from the outer regions of the cluster appears regular
(Buote \& Tsai 1996), detailed \c\ data showed that the central core
is not relaxed (Ettori \etal 2002; Markevitch \etal 2001). In
particular, \c\ revealed a thin, $\sim40$~arcsec-long\footnote{
$\sim40$~arcsec corresponds to a projected length of $\sim46\kpc$,
using an assumed cosmology of $H_0=71$\kmpspMpc, $\Omega_m=0.3$ and
$\Omega_\Lambda=0.7$.} filament of X-ray emission extending away from
the central cluster galaxy to the SSE (Fabian \etal 2001). It is here
in the filament, many kiloparsec away from the influence of the
central radio source, that a more direct connection between the
different components of a cool intracluster medium can be observed.

The X-ray structure coincides spatially with a luminous optical
line-emitting filament discovered two decades earlier by Cowie \etal
(1983; Fig~\ref{fig:images}). The X-ray filament is seen only at soft
energies (0.3-3\keV), with a total $0.5-7$\keV\ luminosity of
$\sim2\times10^{42}$\ergps, barely an order of magnitude more than the
luminosity it emits in the \ha$+$[NII] emission lines. Although clumpy
structure is seen both at optical and X-ray wavelengths, it does not
correspond well spatially.  The filament also shares the general sense
of faint linear structures seen in the $U$-band (McNamara \etal 1996;
Fig~\ref{fig:images}; although note that this filter may also contain
some [OII] line emission), which encompass a bright clump of blue continuum
halfway down the filament, at $\sim21$~arcsec SSE of the central
cluster galaxy. This clump is not obviously associated with a
galaxy in the cluster, and is also seen in the UV images of Mittaz et
al (2001; Fig~\ref{fig:images}).  The colours of this region are very
blue, characteristic of very recent star formation. 

A likely origin for the X-ray/optical/$U$-band filament is
that it is a \lq cooling wake', produced as the host cluster gas cools
around a moving galaxy (Fabian \etal 2001).  The dominant cluster 
galaxy of A\,1795 travels through the cluster gravitational potential
with a peculiar velocity of $+150$\kmps relative to the mean of all
the other cluster galaxies, and at $+374$\kmps faster than those
within the central 270kpc (Oegerle \& Hill 1994).  The most rapidly
cooling gas in the inner intracluster medium would then be
gravitationally focussed onto a line trailing the central cluster
galaxy's direction of motion. This view is supported by the facts that
the radiative cooling time of the X-ray gas is comparable to a
dynamical timescale for the optical filament estimated from its size
and velocity; also that Hu \etal (1985) find that most of the optical
filament shares the velocity of the cluster rather than that of the
central galaxy at its head. The dominant galaxy of A\,1795 also has a
cD envelope extended out to radii of up to 3 arcmin, which is
elongated on the same approximate north-south axis as the filament
(Johnstone, Naylor \& Fabian 1991).  The large-scale distribution of
the galaxies and the X-ray emission from the cluster also both appear
to be elongated in the same direction.

In this paper we present a deeper X-ray image of the core of the
cluster, and optical spectroscopy of the A\,1795 filament, in order to
re-examine the relationship between the different wavelength
components.  This filament is a useful illustration of cooling
behaviour within an intracluster medium, as it well removed from the
complicating effects of a powerful radio source.  We also examine the
kinematic interaction of the ionized gas close to the radio source
near the nucleus, as well as the nature of the massive young star
formation taking place both here, and within the filament. 

\section{X-ray data}

\begin{table}
\begin{tabular}{lllll}
Sequence & Instrument & Exposure & Date of & Type of \\
number   &            & (ksec)   & Observation & Observation \\
790078 & ACIS-S & 14.4 & 2002-06-10 & Cal \\
800001 & ACIS-S & 19.6 & 2000-03-21 & GO \\
800002 & ACIS-S & 19.5 & 1999-12-20 & GO \\
890027 & ACIS-S & 14.3 & 2004-01-14 & Cal \\
890026 & ACIS-S & 14.4 & 2004-01-14 & Cal \\ %
290028 & ACIS-S & 14.6 & 2004-01-16 & Cal \\ %
890029 & ACIS-I & 15.0 & 2004-01-18 & Cal \\
890030 & ACIS-I & 14.9 & 2004-01-23 & Cal \\ %
\end{tabular}
\caption{ \label{tab:xrayobs}
The \c\ X-ray observations used.}
\end{table}

\subsection{Observations and reduction } 

The X-ray filament was first discovered in the $\sim40$ ksec \c\ image
of Fabian et al (2001). We have merged the two original GO datasets of
these observations (sequence numbers 800001/2) with a further six
calibration datasets from the \c\ archive; the observations used are
listed in Table~\ref{tab:xrayobs}.  Three of the calibration datasets
have the core of the cluster placed away from the aimpoint of the
chip; whilst this may introduce artificial structures in the much
larger-scale cluster emission, the features and structure in the
filament itself are unaffected (we have checked this by repeating the
analysis omitting the three offending datasets).  The secondary event
files for each of the datasets were added together with the CIAO
\textsc{merge\_all} contributed script to make a total merged event
file.  The resulting total exposure time for this merged observation
is 126.7~ksec. We extracted an image in the 0.3 to 7\keV\ band from
this merged event file, using the native 0.492 arcsec binning. 

We also obtained spectral information on the filament from a separate
reduction using only the six ACIS-S datasets (ie omitting the two ACIS-I
datasets); these were reprocessed with the latest appropriate gain files.
Time periods from the observations showing flares in the 2.5 to 7 keV
band on the ACIS-S5 CCD (or the S3 if there was no S5 data in the
observation) were removed. This yielded a total ACIS-S observation
time of around 90~ksec in the central part of the filament, and  a total
image was created in the 0.5 to 7~keV band. We smoothed this image
with the accumulative smoothing algorithm (see Sanders in
preparation), which uses a circular top hat smoothing kernel that
grows until the signal to noise is sufficient within the kernel. In
this case we smoothed the image with a signal to noise ratio of 20. 

We created bins on the raw ACIS-S image which contained a signal to
noise of 32 ($\sim 1000$ counts), using a contour binning algorithm
(Sanders in preparation). The routine grows regions from the brightest
pixel on the smoothed map, adding neighbouring pixels nearest in value
from the smoothed map, until the threshold signal to noise is reached.
Pixels lying outside a radius of 1.8 times the radius of a circle with
the same area as the bin were not added, ensuring the regions did not
become very elongated.  We start binning again from the next highest
flux pixel in the image, until all the pixels are exhausted. Any stray
regions are gathered up into the neighbouring bin with the nearest flux in
the smoothed map. Spectra from each of the regions were extracted from
each of the datasets, as well as from corresponding blank-sky fields.
We also calculated responses and ancillary responses for each of the
datasets and regions. The spectra for a bin were added together
(ignoring those datasets with no data for the bin). We added the
responses and ancillary responses together for the region, weighting
them according to the number of counts in the corresponding spectrum.
We added the background spectra together so that the effective
exposure of each component was the same fraction of the total
background, as the number of counts of the foreground spectrum was of
the total foreground spectrum. This procedure is appropriate if the
spectral response of the detector does not change greatly between the
observations. We tested the procedure using faked data with a variety
of different parameters. It produced no systematic temperature or
abundance changes, but produced a slight offset in $N_{\mathrm{H}}$
($\sim 5\times 10^{20}$ atoms~cm$^{-2}$), presumably due to the change
in effective area caused by the build up of contaminant on the ACIS
detector.

We fitted the spectra within \textsc{XPEC} by a \textsc{mekal} model
(Mewe, Gronenschild \& van den Oord 1985; Liedahl, Osterheld \&
Goldstein 1995) absorbed by a \textsc{phabs} model (Balucinska-Church
\& McCammon 1992). In the fits, the temperature, abundance (relative
to Solar), normalisation and absorption were free. We minimised the
C-statistic (Cash 1979) in the fits. 

\subsection{Results} 
\begin{figure}
\begin{centering}
       \psfig{figure=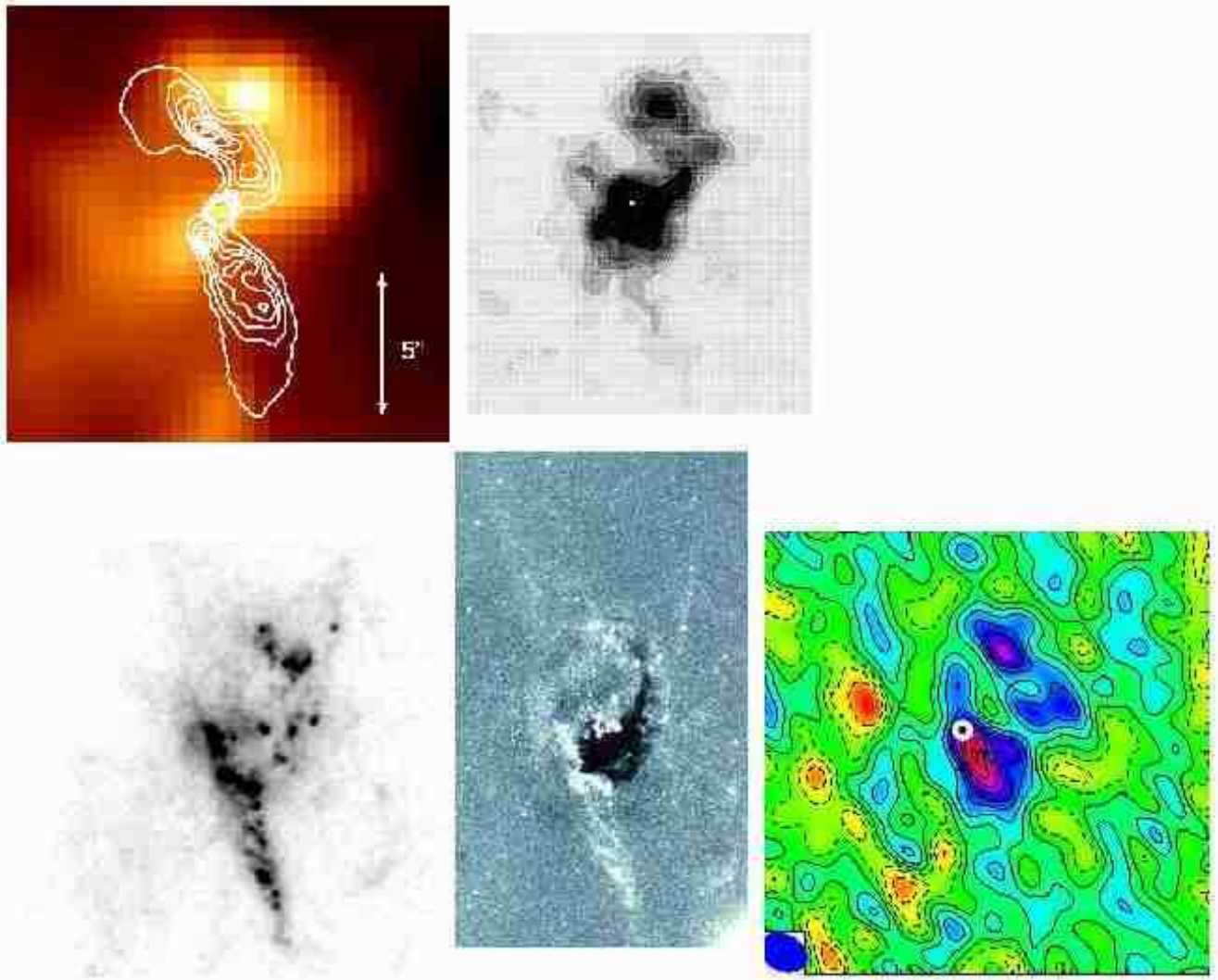,width=0.99\textwidth,angle=0}
\end{centering}
\caption{\label{fig:zoomimages}
The environment local to the radio source 4C26.42 in the central
galaxy of A\,1795 in different wavebands. The images are matched in
scale, and offset in the vertical direction so that the position of
the nucleus in each image is aligned horizontally in each row. 
Top row: \newline
Left) the smoothed \c\ 0.3-7\keV\ X-ray image, with white contours of radio emission
at 3.6cm superposed (courtesy of G. Taylor). The image size is 
$16\times16$ arcsec. \newline
Right) the \ha+[NII] optical line emission (adapted from van Breugel et al
1984), with the position of the radio nucleus marked by a white
dot. The image is approximately $12\times14$ arcsec. \newline
Lower row: \newline
Left) the far-UV continuum and Ly-$\alpha$ line emission from the HST
STIS (F25QTZ and F25SRF2 filters) taken from O'Dea et al (2004). The
image is approximately 16 arcsec square. \newline
Middle) galaxy-subtracted HST image (F702W$+$F555W filters) taken 
from Pinkney et al (1996). The dust lane absorption shows as black, and 
emission features as white. The scale is  10.5$\times$18 arcsec, and
the position of the nucleus is marked by a red \lq N'.  \newline
Right) the continuum-subtracted CO(2-1) emission from Salom\'e
\& Combes (2004). The position of the radio nucleus is marked by a
white dot and the beamsize is indicated by the solid oval in the lower
left-hand corner of the image. The image is 16 arcsec square.
\newline
}\end{figure}

\begin{figure}
       \psfig{figure=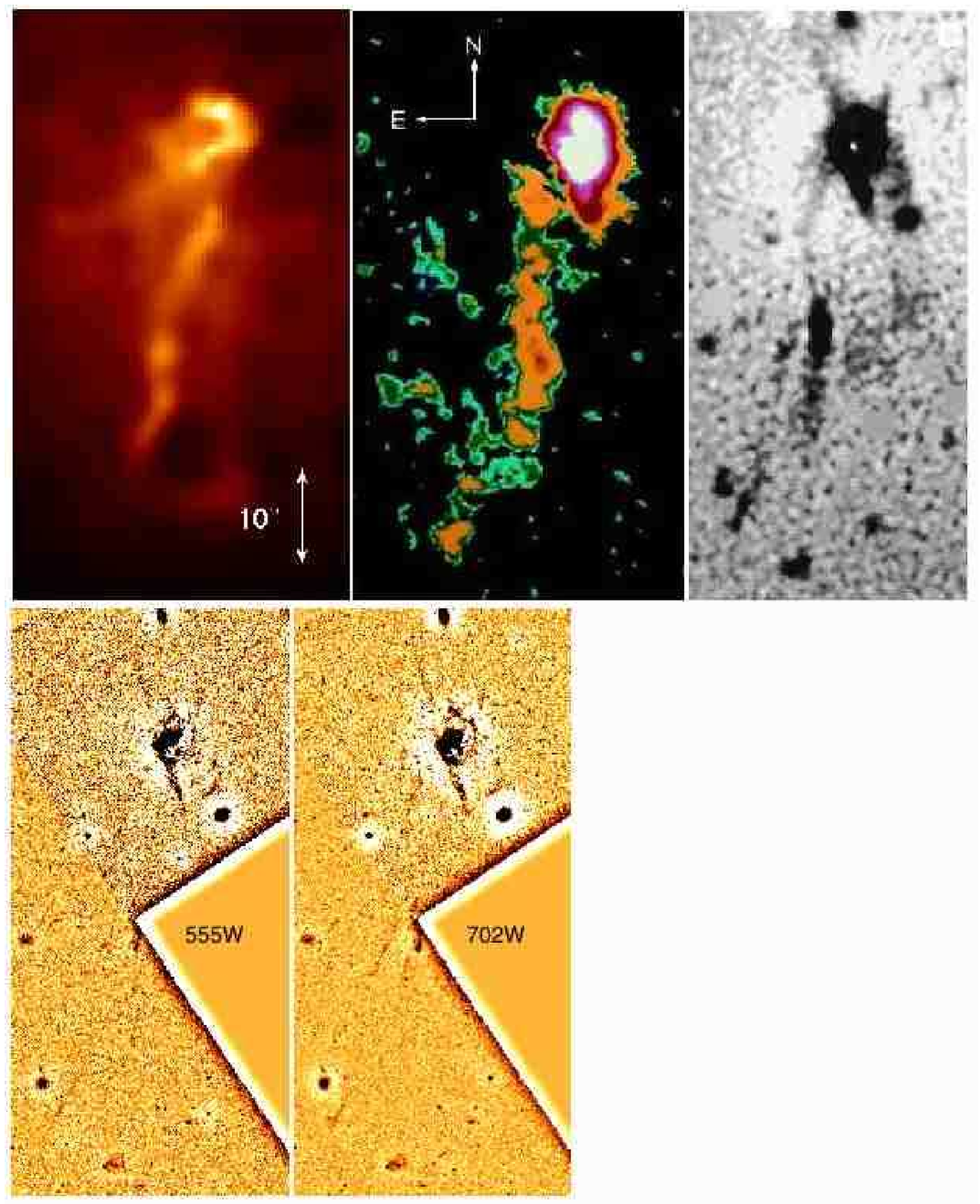,width=0.99\textwidth,angle=0}
\caption{ \label{fig:images}
{\bf Top} (from left to right): \newline
Adaptively-smoothed soft \c\ X-ray (0.4-7\keV) image of the filament 
south of the central galaxy in A\,1795; the image is 
$37\times65$ arcsec. \newline
The filament in the light of the 
\ha+[NII] line emission from Cowie \etal (1983); 
 the image is $\sim37\times$65 arcsec. \newline 
The $U$-band continuum image (with the galaxy
continuum subtracted) adapted from McNamara \etal (1996).  The image
is 30$\times$65 arcsec. The nucleus is shown by a white dot. \newline
The images are to scale, and
the position of the radio nucleus in each is aligned horizontally in
each row.\newline
{\bf Bottom}: 
Unsharp-masked WFPC2 HST images of the filament of A\,1795 through the F555W (left;
corresponding to virtually line-free emission) 
and the  F702W (right; which includes the redshifted \ha$+$[NII]
emission lines) filters.
Each image covers an area of 
 $30\times65$ arcsec. \newline 
}
\end{figure}

\begin{figure}
     \psfig{figure=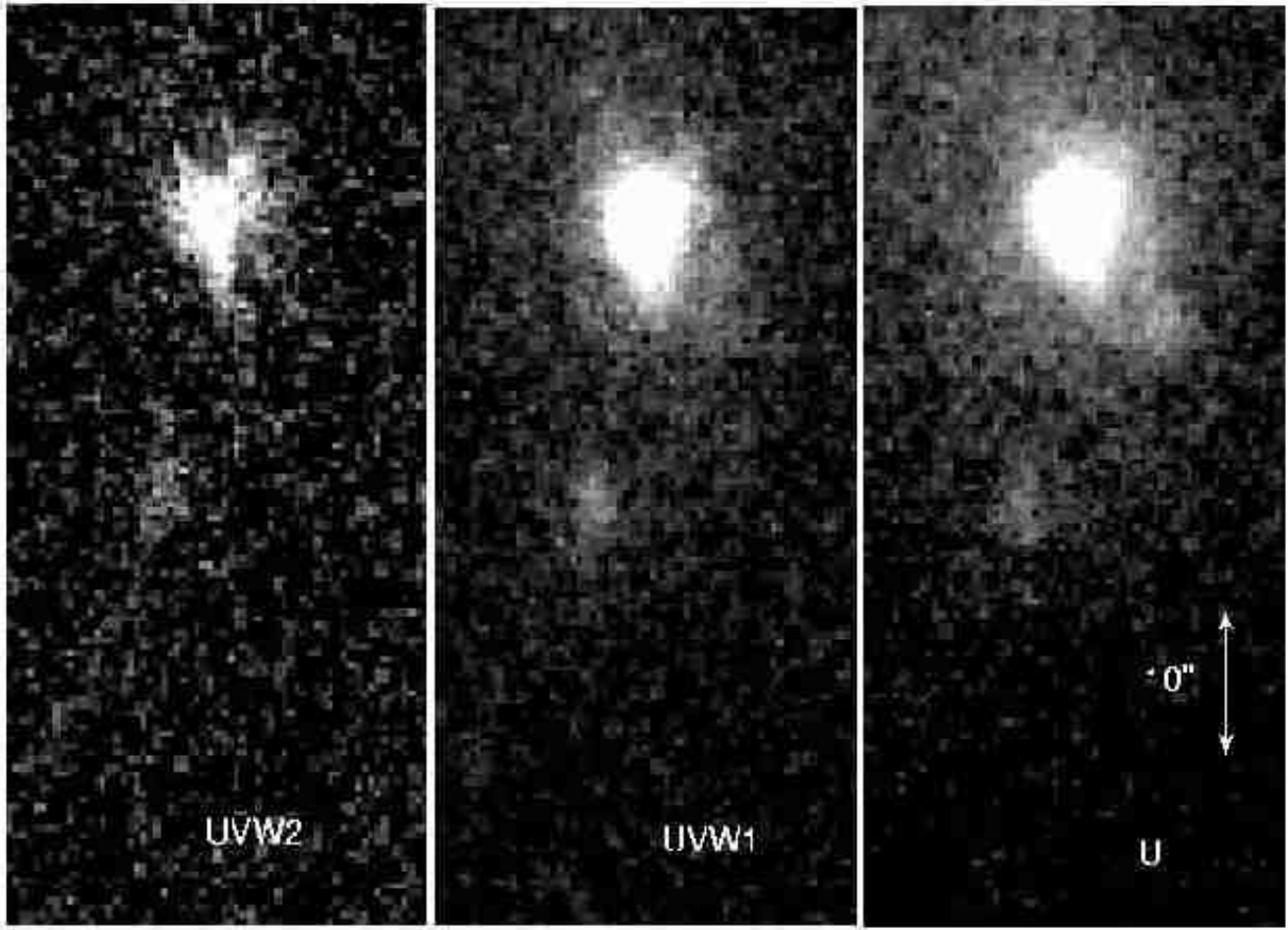,width=0.99\textwidth}
\caption{ \label{fig:uvimages}
{\sl XMM-Newton} optical monitor images of the central galaxy of A\,1795
in the light of the (from left to right) $UVW2$ (1800-2250\AA;
corresponding to around 1690-2115\AA\ at the redshift of the galaxy),
$UVW1$ (2450-3200\AA; corresponding to 2305-3010\AA) and $U$
(3000-3900\AA; corresponding to 2820-3670\AA) bands. 
Each image covers an area of 
 $30\times65$ arcsec. \newline 
}
\end{figure}

The final 0.3-7\keV\ merged image is shown (on two different spatial
scales) in Figs~\ref{fig:zoomimages} and \ref{fig:images}, along with
images in other wavebands on a matched spatial scale for direct
comparison: individual images in these figures are adapted from those
published by O'Dea et al (2004), Salom\'e \& Combes (2004), McNamara
et al (1996), Pinkney et al (1996), van Bruegel et al (1984) and Cowie
et al (1983); the 3.6cm radio data are courtesy of G. Taylor (from Ge
\& Owen 1993); the {\sl XMM-Newton} UV data of Mittaz et al (2001)
are shown in Fig~\ref{fig:uvimages}. The X-ray images are shown after adaptive
smoothing using the \textsc{asmooth} algorithm with a minimum
significance of 3$\sigma$. 

X-ray emission is now evident at the position of the radio nucleus, at
RA 13:48:52.44 Dec +26:35:34.5 (J2000), with a 1-7~keV luminosity of
$1.4\times10^{40}$\ergps (assuming a power-law spectrum with
$\Gamma=2$ and only Galactic line-of-sight absorption). The brightest
peak of X-ray emission, however, lies 4.5 arcsec to the NNW of the
nucleus, spatially coincident with the bright peaks seen in both the
CO(2-1) and \ha+[NII] line emission images (Fig~\ref{fig:zoomimages}).
The emission in this region shares the same morphology in the three
wavebands, appearing to curve round the outside of the northern radio
lobe, first to the west then back round north-east to the offset
bright peak. There is a prominent gap in the X-ray emission south of
the active nucleus and before the start of the filament, at the
location of the southern radio lobe. 

The new X-ray image refines the clumpy structure seen by Fabian et al
(2001) within the X-ray filament. As noted in that paper, although the
sense of the filament is the same in the X-ray, \ha+[NII] and UV
wavebands, there is no detailed one-to-one correspondence of
features, particularly within the northern half of the filament. A
bright X-ray blob 22.5 arcsec south of the nucleus coincides well with
the brightest peak in the \ha+[NII] filament, but lies a couple of
arcseconds to the south of the prominent UV continuum blob. In each
band this bright peak is accompanied by a second, fainter knot around
5 arcseconds due south.  There is no sign of any soft X-ray emission
associated with the two much fainter cross-spurs seen in the optical
emission-line image, running into the filament from the north-east. 

The top panels of Figs~\ref{fig:slitplots} and \ref{fig:slitplotszoom}
show the X-ray flux intensity along a 2~arcsec cut across the nucleus
and filament, and a parallel cut 3 arcsec to the east (these two cuts
correspond to the optical slit positions detailed in the next
section, also see Fig~\ref{fig:botho2}). The same data is displayed on two
different scales to distinguish the general properties of the filament
(in Fig~\ref{fig:slitplots}) and the properties very local, and in
relation to, the radio source (in Fig~\ref{fig:slitplotszoom}).  

\begin{figure}
\psfig{figure=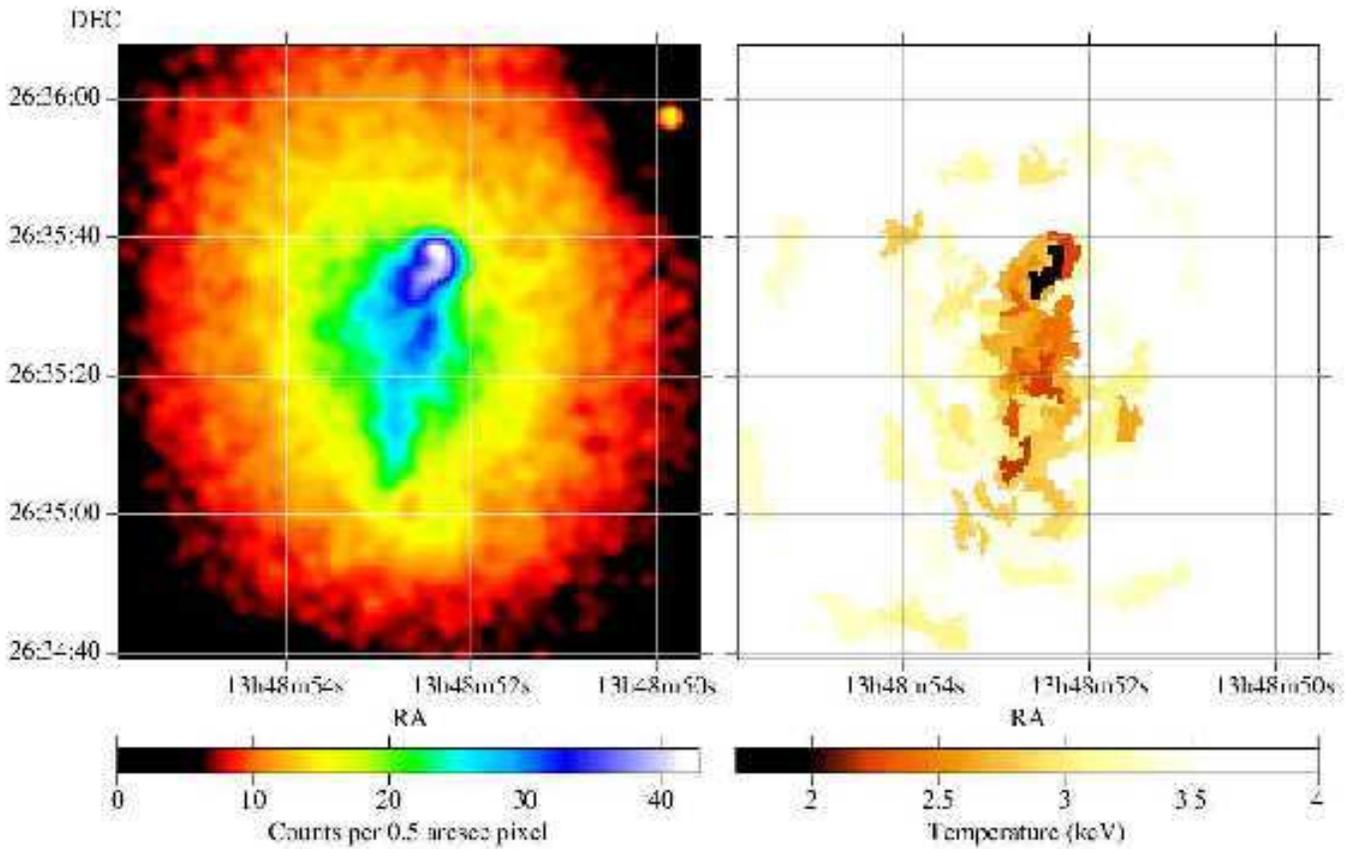,width=1.0\textwidth}
\caption{\label{fig:ximages} 
(Left) The number of X-ray counts (0.5-7\keV) within a 0.5 arcsec pixel for the
six summed \c\ ACIS-S datasets for the core of the 
A\,1795 cluster of galaxies, showing the filamentary structure.
The image has been smoothed with a Gaussian of 1~arcsec.   \newline
(Right) A map showing the variation in X-ray temperature along the
filament, created from 
spectral fitting. }
\end{figure}

Fig~\ref{fig:ximages} shows the number of counts for the combined six
ACIS-S datasets in 0.5 arcsec pixels and the 0.5-7\keV\ energy range
(smoothed with a Gaussian of two pixels, corresponding to 1~arcsec;
left-hand panel), and the projected emission-weighted temperature of
each region (right-hand panel). The uncertainties on each individual
temperature measurement vary from $\sim 8$~per~cent along the filament
out to $20$~per~cent at the edge of the region shown. The central
region around the nucleus now shows emission down to 1.8~keV. The
filament is seen to be at 2.1-2.6~keV, substantially cooler than that
of the surrounding intracluster medium, with the coolest regions (at
$\leq2$\keV) located in the brightest structures around the radio
source (Fig~\ref{fig:ximages} -- the bright patch of emission just 2
arcsec south-east of the nucleus, and the peak in the X-ray emission
to the north-west). The components {\sl within} the filament are, in
comparison, not directly coincident with the brightest regions of
X-ray emission. 

\begin{figure} 
\psfig{figure=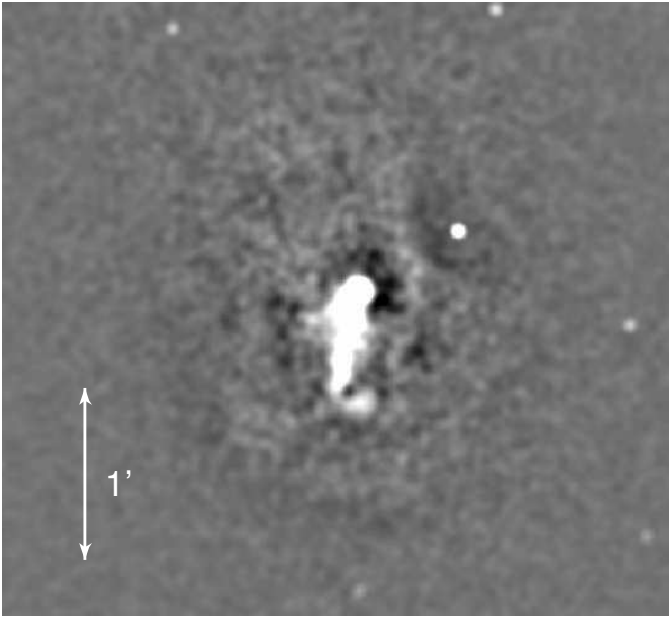,width=0.5\textwidth}
\caption{\label{fig:xcircle} Unsharp-masked 0.4-7\keV\ X-ray image of
the A\,1795 cluster of galaxies, showing the depression with a
circular edge that is centred around the bright X-ray source lying 42
arcsec to the NW of the nucleus.  The image has been smoothed by
1.5~arcsec, and spans 214 arcsec from north to south. } 
\end{figure}

A hole in the X-ray emission is apparent to the West of the end of the
X-ray filament in Fig~\ref{fig:ximages}, centred at RA~13:48:52.7,
Dec~26:35:01.8. This is not an artefact of adaptive smoothing, as it
is also present in the original unbinned data.  No object is visible
at this position in the DSS. A further structure is seen as a
semi-circular depression in the X-ray emission, approximately centred
on the bright point source (at RA 13:48:49.9, Dec 26:35:57.5),
$\sim42$ arcsec to the NW of the nucleus. This depression is again a
feature visible in the raw data, but becomes more apparent after
unsharp-masking the image (Fig~\ref{fig:xcircle}); \ie the dataset was
subtracted by itself smoothed by 10~arcsec. The deficit in 
X-ray emission does not seem to be due to X-ray absorption, as the
spectral image shows no obvious excess of N$_{\rm H}$ is in this
region.  It could represent an infalling cluster, but no major
galaxies are seen in this region, apart from a faint source in the HST
data coincident with the central X-ray point source.  The structure
may be a ghost radio bubble as seen, for example in MKW3s (Mazzotta et
al 2002), but there is no low-frequency radio emission coincident with
this region, and it would be unusually small in size, with a radius of
15.5 arcsec (corresponding to $\sim17.8$\kpc at the redshift of
A\,175). 

\onecolumn
\twocolumn

\onecolumn

\twocolumn

\onecolumn
\twocolumn

\onecolumn
\twocolumn

\section{HST images}
\subsection{Archival observations}
High spatial-resolution HST images of the central cluster galaxy in
A\,1795 have been published in many papers (eg O'Dea et al 2004;
Pinkney et al 1996; McNamara et al 1996). We extracted two archival
WFPC2 HST observations where some of the filament was also in the
field of view.  These two datasets each have a total observation time
of   1780~sec, and are taken through the F555W (U2630404T $+$ U2630405T
$+$ U2630406T) and F702W filters (U2630401T $+$ U2630402T $+$
U2630403T). The F555W (which is a $V$-band filter) observation is
dominated by stellar continuum emission at the redshift of A\,1795,
the only line emission being the comparatively weak H$\beta+$[OIII].
The F702W observation includes the strong redshifted emission lines of
\ha$+$[NII] in its bandpass. 
The raw data were combined using standard STSDAS 
software within IRAF. In order to show up the faintest structures
within the filament, we unsharp-masked the data by subtracting the
data by itself smoothed by 20 pixels. The final images
were then smoothed by 1 pixel, and the resulting images are shown in
Fig~\ref{fig:images}. 

\subsection{Results} 
\begin{figure}
\psfig{figure=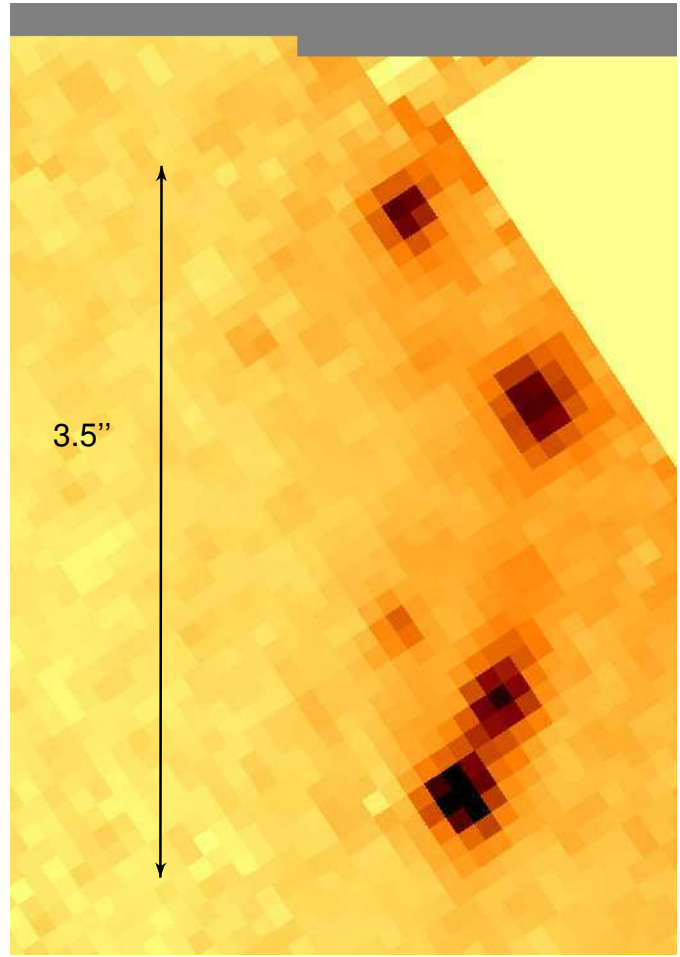,width=0.5\textwidth}
\caption{ \label{fig:knots}
A close-up of the continuum knots within the filament 
from the unsharp-masked 702W HST image shown in Figure~\ref{fig:images}. }
\end{figure}

The images in both HST bands show linear
structures emerging from the galaxy to the NW and NE, and the
continuum forming the star formation around the southern radio lobe.
In addition, the F702W image shows very faint, thin and double linear
structures to the SSE, echoing the filament as seen in the
ground-based $U$-band image of McNamara et al (1996;
Fig~\ref{fig:images}). The fact that these filaments are seen only in
the F702W and not the F555W band strong suggest that they are due to
redshifted line emission of \ha\ and [NII] in the filter bandpass. 
Unfortunately much of the star forming region half-way along the
filament is located just at the join between the PC and WF chips. A
chain of at least five resolved knots are visible in this location,
however, and are shown in close-up in Fig~\ref{fig:knots}. 
Each knot is approximately 0.3~arcsec across, corresponding to a size
of around 170~parsecs at the redshift of the galaxy, and they are spread
over a length of 3.5~arcsec ($\sim$4~kpc). 
All five knots are visible in both the F555W and F702W
bands, strongly confirming that they are primarily continuum in
origin. We do not attempt photometry of the knots as they are so close
to the edges of the chips, rendering the results unreliable. 

\section{Optical spectroscopy data}

\subsection{Observations and Reduction } 
\begin{figure}
\psfig{figure=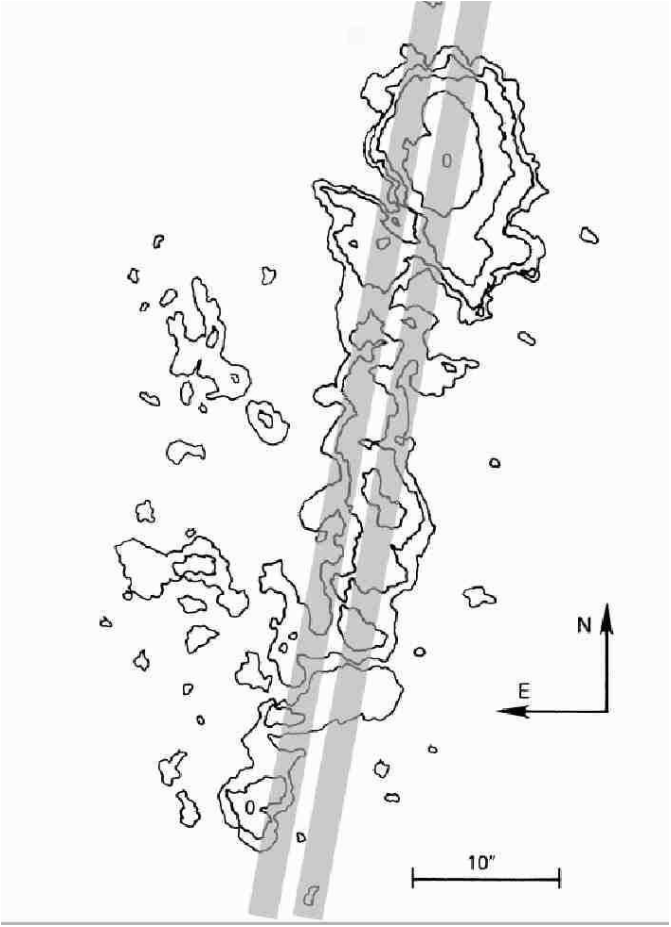,width=0.5\textwidth,angle=0}
\caption{\label{fig:slitpos}
The positions of the two slit observations (grey boxes) shown against
contours of the \ha+[NII] line emission from the filament (adapted
from Hu, Cowie \& Wang 1985). }\end{figure}

Optical observations were
taken on 2001 May 11 and on 2002 February 03, using the ISIS
spectrograph at the 4.2m William Herschel Telescope on La Palma.  On
each occasion the optical filament was observed through a
2~arcsec-wide slit oriented at a position angle of 170\deg for a total
exposure time of 5400~s. The slit for the observation in 2001 was
positioned to cover the centre of the galaxy and the bulk of the
filament, including the blue knots of star formation
(Fig~\ref{fig:slitpos}). The slit position for the 2002 observation
was at the same position angle but offset by 3 arcsec to the East
(Fig~\ref{fig:slitpos}), in order to cover the eastern side of the
filament not encompassed by the first slit position. The seeing during
both nights was 1--1.2~arcsec, and there was a small amount of (grey)
dust during the 2002 observation. The light from the slit was split
with the 5700 dichroic (with the GG495 blocking filter) into the blue
and red arms of the spectrograph, where it was dispersed by the R158B
and R158R gratings onto the EEV12 (blue) and TEK4 (red) CCDs, yielding
wavelength dispersions of 1.62 and 2.90\AA/pixel respectively.  The
galaxy was observed at airmasses ranging between 1.050 -- 1.002, so
the effects of atmospheric dispersion are minimized.

The red and blue datasets were each reduced using standard software
packages within IRAF. The data were first bias-subtracted and
flattened using long exposures of tungsten lamps; the individual
spectra were corrected for a slight atmospheric dispersion before
combining, after which they were wavelength-calibrated, sky-subtracted
and flux-calibrated before the final correction for atmospheric
extinction and Galactic reddening (A$_{\rm V}\sim0.080 $, assuming
$N_{\rm H}=1.2\times10^{20}$\pcmsq, from Stark et al 1992). The final
two-dimensional spectra have fluxed wavelength ranges of
3200--5400\AA\ and 6325--9300\AA, and spatial binnings of 0.2 and 0.36
arcsec per pixel in the blue and red arms respectively. The data were
de-redshifted using a redshift of $z=0.0633$.

It is common for central cluster galaxies with strong line emission to
also have some level of intrinsic dust emission (E(B-V)$\sim0.3$), as
determined by the Balmer decrement between \ha\ and \hb\ (Allen 1995;
Crawford \etal 1999). In particular, the dominant cluster galaxy of
A\,1795 shows a patchy dust lane across its central regions in the HST
images (Fig~\ref{fig:zoomimages}; Pinkney et al 1996; McNamara et al 1996) and the Balmer
decrement measured from a long slit spectrum of the whole galaxy
indicates that E(B-V)$\le0.3$ (Allen 1995; Crawford \etal 1999). In
our current spectra, the \ha\ and \hb\ emission lines are detected on
different arms of ISIS, and there are systematic uncertainties in
matching the two datasets. We are also wary of applying a blanket
correction from an averaged spectrum, given the observed patchy
structure of the dust lane. We thus do not correct the spectra for the
intrinsic reddening, but assess later any effect this may have in
understanding the line emission properties and the continuum
contribution from massive young stars.

\subsection{Analysis and Results }

For all the spectral results, we shall refer to distances along
the slit relative to the measured peak of the galaxy continuum. For
the spectra from the blue arm, this peak is assessed from the profile
of only the very reddest continuum available. In the central slit
position we take this continuum peak to be coincident with the core of
the radio source. We adopt the convention that negative distances lie
to the south of the nucleus, and positive distances to the north.

\begin{figure}
\hbox{
\psfig{figure=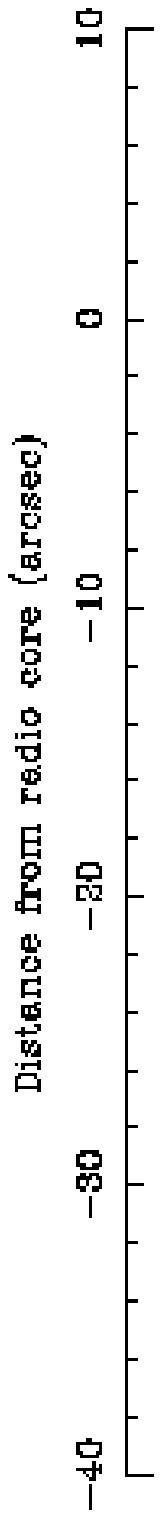,width=0.1056\textwidth}
\psfig{figure=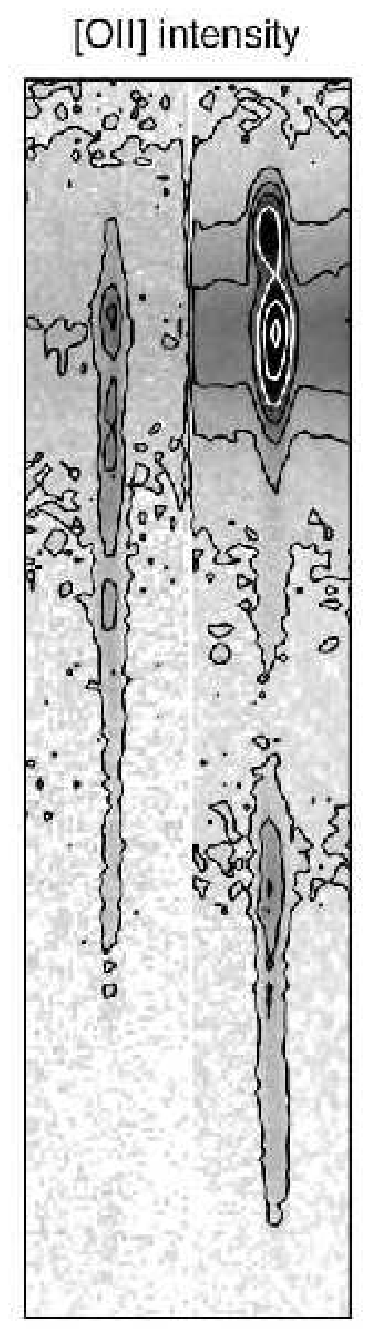,width=0.3\textwidth}
}
\caption{\label{fig:botho2}
Two-dimensional spectra of the emission from the [OII] doublet in both
slits; the slit centred on the galaxy is shown to the right, and
that offset to the east shown on the left. The images each span
$\pm$50\AA\ from the peak of the [OII] line emission 
($x$-axis), and $+10$ arcsec to
the north and $-40$~arcsec to the south of the galaxy nucleus ($y$-axis). 
The intensity greyscale is from 0 (white) to
3$\times10^{-17}$\ergpspcmsqpA (black). Black contours are drawn at
0.0008, 0.0030, 0.0065
and 0.0115$\times10^{-15}$\ergpspcmsqpA  and the white contours at
0.02, 0.045 and 0.07$\times10^{-15}$\ergpspcmsqpA. 
The continuum spectrum of the blue
clump  is just visible around 22~arcsec south of the nucleus in the
central slit (right hand panel). }
\end{figure}

The filament is of low ionization, luminous in the lines of
\ha$+$[NII] and [OII]$\lambda$3727, and emitting less strongly in
\hb$+$[OIII], [NI], [SII] and [OI]. The central slit shows a prominent
blueshifted secondary peak of emission to the north of the galaxy, 
and encompasses the brightest parts of the filament south of the
location of the blue clump (which is of course unresolved into the
separate knots in our ground-based spectrum) -- which itself shows up
as a weak continuum in our data (Fig~\ref{fig:botho2}). The data from
the eastern slit do not include much of the secondary peak in emission
to the north, but do \lq fill in' the upper part of the filament
between the blue blob and the dominant galaxy (Fig~\ref{fig:botho2}). 

\begin{figure}
\vbox{
\psfig{figure=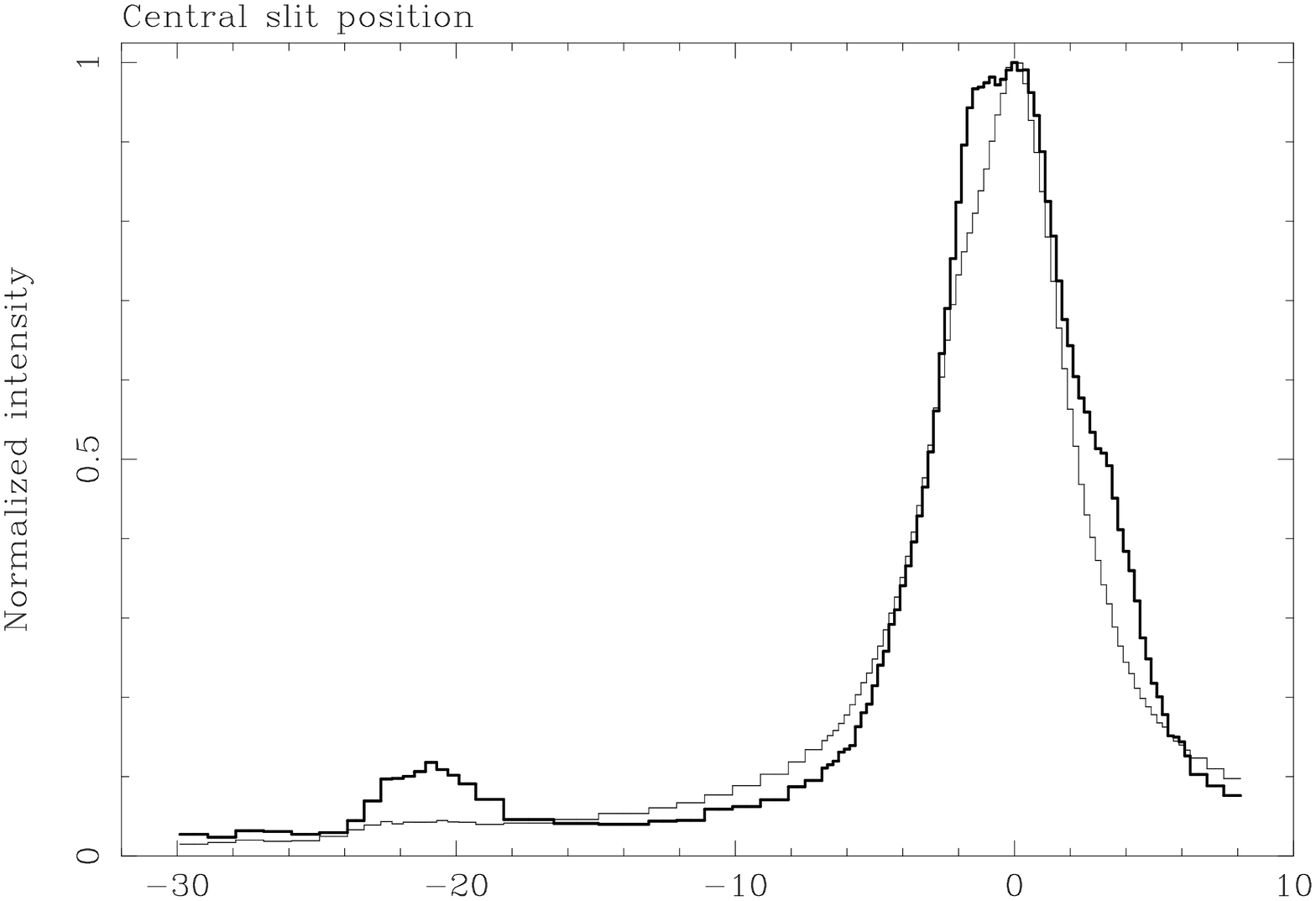,angle=270,width=0.5\textwidth}
\psfig{figure=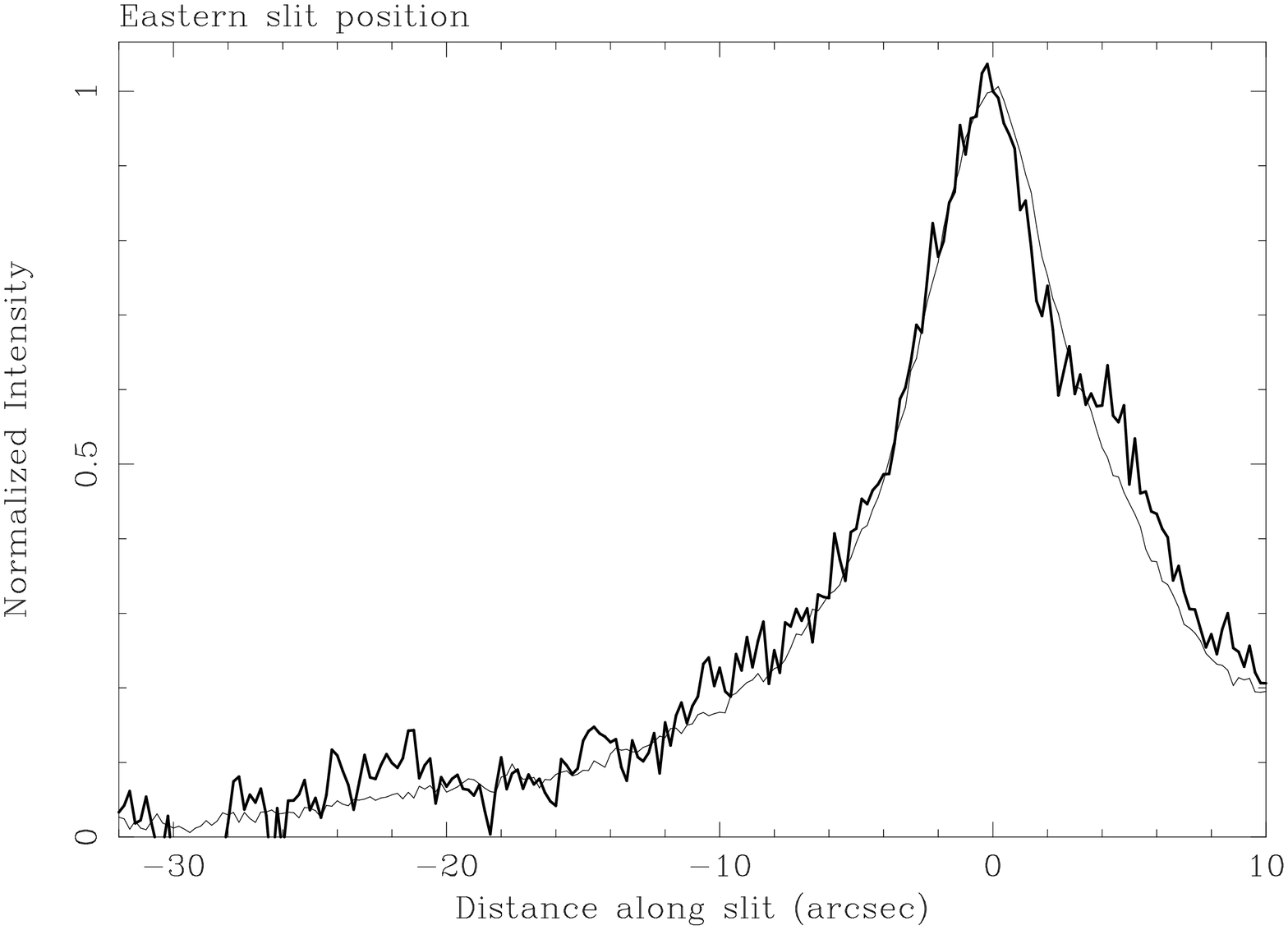,angle=270,width=0.5\textwidth}}
\caption{\label{fig:cuts}
  The spatial intensity profile of the continuum along the slit
  centred on the galaxy nucleus (top), and along the slit to the east
  (bottom); both in a blue (rest-frame 3200-3650\AA; thick line) and a
  red passband (4500-4800\AA; thin line). Both profiles have been
  normalized at the galaxy centre. Positive distances are to the north
  of the radio core, negative distances to the south. The spectra from
  the central slit position show three clear excesses of blue
  continuum centred at 3.8 arcsec to the north of the galaxy core,
  around 1.5 arcsec to the south, and at around 21 arcsec south, in
  the filament. The spectra from the eastern slit shows slight
  excesses at 22 arcsec south of the galaxy, and around 4 arcsec to
  the north. }
\end{figure}

In Fig~\ref{fig:cuts} we show the spatial variation of continuum
intensity along each slit position in the blue in for two line-free
wavebands: 3200-3650\AA\ and 4500-4800\AA\ (in the rest-frame).  
There are three main concentrations of excess blue continuum; one
centred at about 3.8~arcsec to the north of the galaxy core, one at
around 1.5~arcsec to the south (these are the \lq lobes' of blue light
lying at the boundaries of the central radio source) and that from the
blue clump lying halfway along the emission-line filament at
$\sim$21~arcsec south.  This latter blue continuum feature appears to
have a fairly sharp edges.  The spectra from the eastern slit show
much weaker contrasts in continuum colour, but still catch the edges
of the blue structures 4 arcsec to the north and 22 arcsec to the
south of the galaxy. The ratio of the fluxes in these bands (blue/red)
is also included in the second panel from the top in both
Figs~\ref{fig:slitplots} and \ref{fig:slitplotszoom}. 

\subsubsection{Emission-line fitting method}
\label{sec:linesfit}
The [OII]$\lambda3727$ doublet was fit as single Gaussian (which is an
adequate fit at this spectral resolution) using QDP (Tennant 1991),
with the continuum represented by a constant ($+$linear only where
necessary) function over the (rest) wavelength interval
$3550-3780$\AA. We fit other emission lines together in close sets
also using QDP, with [OIII]$+$\hb\ fit together, and
\ha$+$[NII]$+$[SII]$\lambda\lambda$6717,6731$+$[OI]$\lambda\lambda$6300,6363
as another set. Within each such fitting set, the individual emission
lines are modelled as Gaussian fixed to be at the same redshift and
velocity width as each other, with the doublets of [NII] and [OIII]
fixed to be in the correct intensity ratios.

To fit the [OIII]$+$\hb\ complex, we first scale (at 4640-4840\AA) and
subtract a galaxy continuum template (created as the mean of
twenty-four non line-emitting central cluster galaxies; taken from
Crawford \etal 1999) to take into account the \hb\ present in
absorption in the underlying galaxy starlight. The lines are then fit
over the (rest) wavelength range of 4700-5050\AA, with no continuum
contribution assumed.  The subtraction of a scaled galaxy template is
not necessary beyond a distance of 30~arcsec south from the galaxy
core, as the light from the galaxy becomes negligible.  The results
from these fits yield our initial results for the emission-line
intensities. There is the possibility, however, that the averaged
central galaxy template may not represent the \hb\ absorption profile
correctly in regions with a prominent excess blue continuum; this is
particularly true if this blue excess is comprised of stars with
prominent Balmer absorption features. This would result in errors in
our estimates of the \hb\ fluxes. We thus later carry out a check by
fitting the [OIII]$+$\hb\ emission-line complex at the same time as a
full spectral modelling of the blue galaxy continuum (see
section~\ref{sec:revisitlines}).

The red data are first corrected for atmospheric absorption, and then
also have a scaled (at 6800-7000\AA) template non-cooling flow galaxy
continuum subtracted before the emission-line fitting, again to
counter the effect of any underlying \ha\ in absorption that may be
present.  Although the \ha\ absorption level in the galaxy continuum
may again be inaccurate in regions showing strong massive star
formation, the effect is expected to be weaker than for the \hb\
emission, as the \ha\ emission flux is so strong; thus we do not
correct for it later.  Where the emission lines show clear evidence
for two velocity components across the centre of the galaxy (see
section~\ref{sec:kinematics}), we also model the \ha+[NII] emission
lines by two independent line complexes, each with its own redshift and
velocity width. 

\subsubsection{Results: Line intensity and intensity ratios} 

The \ha\ line emission is distributed very similarly to the [OII]
 flux, and its variation as a function of distance along the slits is
 shown in the third panel from the top in Figs~\ref{fig:slitplots} and
 \ref{fig:slitplotszoom} (the intensity plots are from fitting the
 line emission with only a single velocity component).  Other lines,
 such as the [OIII] and \hb\ follow the same distribution, albeit at a
 lower signal.  The central slit shows three distinct peaks of the
 emission. One is centred about 3.6~arcsec north of the nucleus,
 comprising the secondary peak to the NW; and the second peak lies
 just less than 0.4~arcsec south. Both these \ha\ peaks are located
 slightly closer to the nucleus than the blue light lobes (compare to
 the distribution of the excess blue light shown in the second panel
 down in this figure). The third brightest region of line emission
 peaks at 23~arcsec south of the nucleus, slightly further from the
 central galaxy than the blue continuum clump.

The line emission from the eastern slit \lq fills in' the northern
half of the filament between the blue clump and the main galaxy (ie
$-$2 to $-$16 arcsec).  It is structured into a strong peak just
0.7~arcsec to the north of the galaxy centre, and three clear clumps
of \ha\ emission between 2 and 8~arcsec south, and another peaked
around 12~arcsec south. A final peak of emission is visible right at
the tail of the filament, at 36~arcsec south of the galaxy. Most of
this structure is apparent in the [OII] image in Fig~\ref{fig:botho2}.
The \ha\ flux from the eastern slit position is around a factor of ten
less than that from the central slit position. This partly because it
is offset from the brightest part of the galaxy, to where the emission
line intensity has greatly dropped, and also partly due to the fact
that these second set of observations were taken through conditions of
(grey) dust.

The intensity ratios of [NII]$\lambda$6584/\ha\ and
[SII]$\lambda$6716/\ha\ are plotted in the fourth panel from the top
in Fig~\ref{fig:slitplots} and Fig~\ref{fig:slitplotszoom}. The
[SII]$\lambda$6716/[SII]$\lambda$6731 ratio is shown in the fifth
panel only of Fig~\ref{fig:slitplotszoom}, as it can be plotted only for
where the [SII] lines are strongest, within $\pm$10~arcsec of the
galaxy core.  We do not plot the line intensity ratios from the blue
data as they should at this stage be treated cautiously for two
reasons: firstly, the \hb\ emission line is the one most likely to
have its intensity affected by the presence of massive young stars;
and secondly because the large wavelength separation between
[OII]$\lambda3727$ and the [OIII]+\hb\ complex means that any
intrinsic reddening may mask or mimic changes in ionization.  We
approach the blue line ratios later as part of the stellar continuum
modelling.  

\subsubsection{Results: Kinematics of the emission-line gas}
\label{sec:kinematics}

\begin{figure}

\psfig{figure=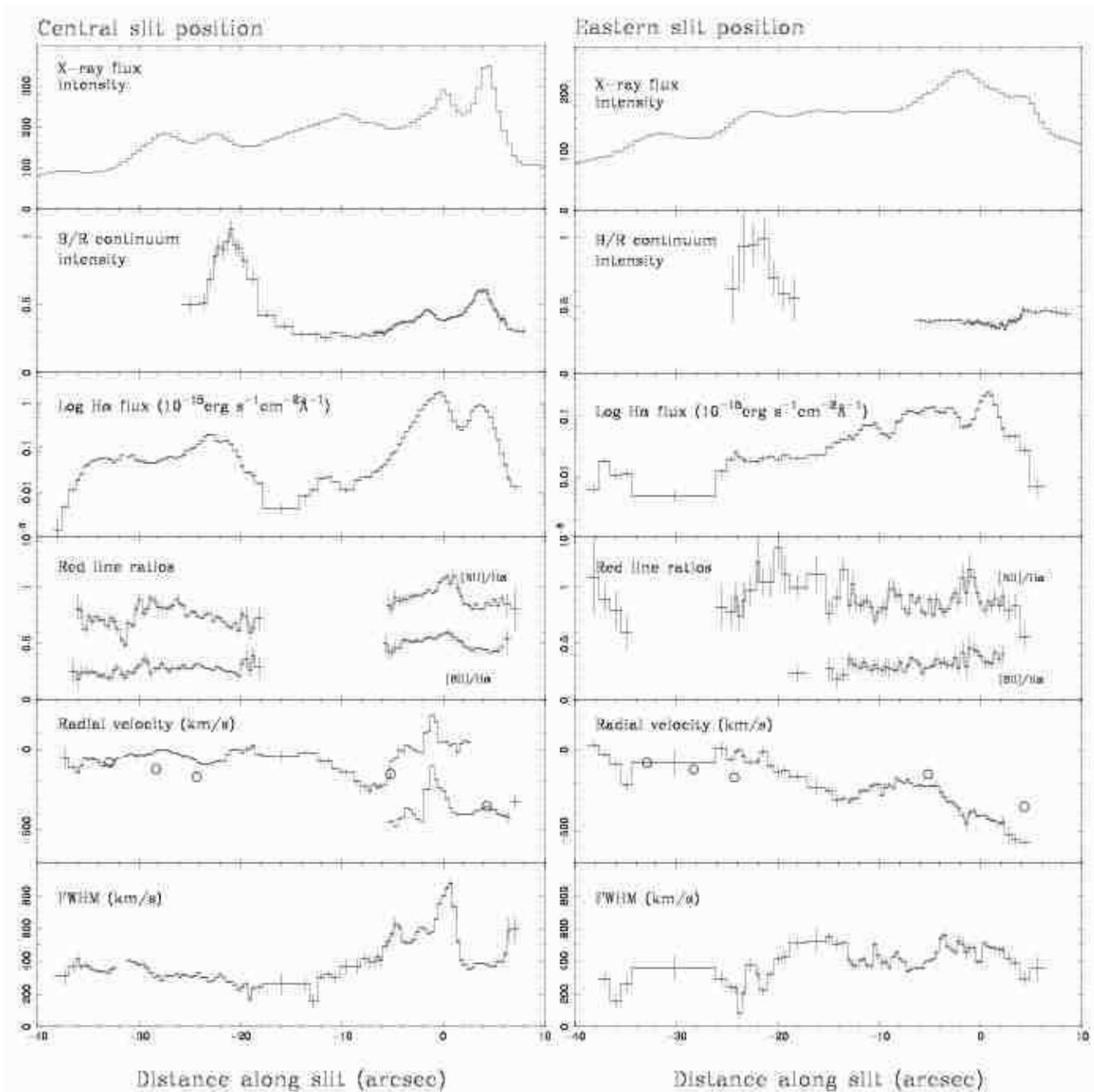,angle=0,width=0.99\textwidth}
\caption{ \label{fig:slitplots}
  The variations in the spectra along the central (left) and eastern
  (right) slit positions. Panels from top to bottom show: \newline 
the  intensity of the X-ray flux along the slit from the
adaptively-smooted image ; 
the ratio of the blue to the red continuum values (in the passbands
given in the caption to Fig~\ref{fig:cuts}); 
the intensity of the \ha\ emission line;
the intensity ratios of [NII]$\lambda6584$/\ha\ and  [SII]$\lambda6717$/\ha; 
the radial velocity of the emission relative to the galaxy nucleus (the large open circles represent values estimated from the kinematic map of Hu \etal 1985); 
and the FWHM of the \ha+[NII] lines (all when modelled by only one velocity component), \newline 
In all figures, positive distances are to the north of the radio core,
negative distances to the south.  The $y$-axis scales differ 
between the left and right panels {\sl only} for the X-ray flux and the \ha\ 
flux intensities.}
\end{figure}

The bulk of the filament is blueshifted by about $-50\pm50$\kmps
compared to the galaxy itself (Fig~\ref{fig:slitplots}; also see Hu et
al 1985), with a sharp decrease to around --200\kmps around 7~arcsec
south of the galaxy. The radial velocity observed in the part of the
filament observed through the eastern slit position is at a slightly
higher blueshift, with a clear blue gradient towards the central
cluster galaxy, where the emission is offset by as much as --400\kmps.
The increased velocity width (Fig~\ref{fig:slitplots}) and clear
double structure seen in the profiles of the line emission spanning
the nucleus (from $-6$ to $+$2.5~arcsec) in the central slit position
indicate that (at least) two velocity components are present here. We
thus also fit this region with two independent \ha+[NII] line emission
complexes in order to get the relative velocity profiles of the two
separate components. The two components are separated by around
400\kmps, and both show a sharp redshift jump of about 200\kmps
between --2~arcsec and the nucleus, and beyond about 3~arcsec north
only one velocity component is apparent, that around --400\kmps. The
northern end of the eastern slit data shows an even more extreme
blueshift of around --600\kmps.

\begin{figure}
   \psfig{figure=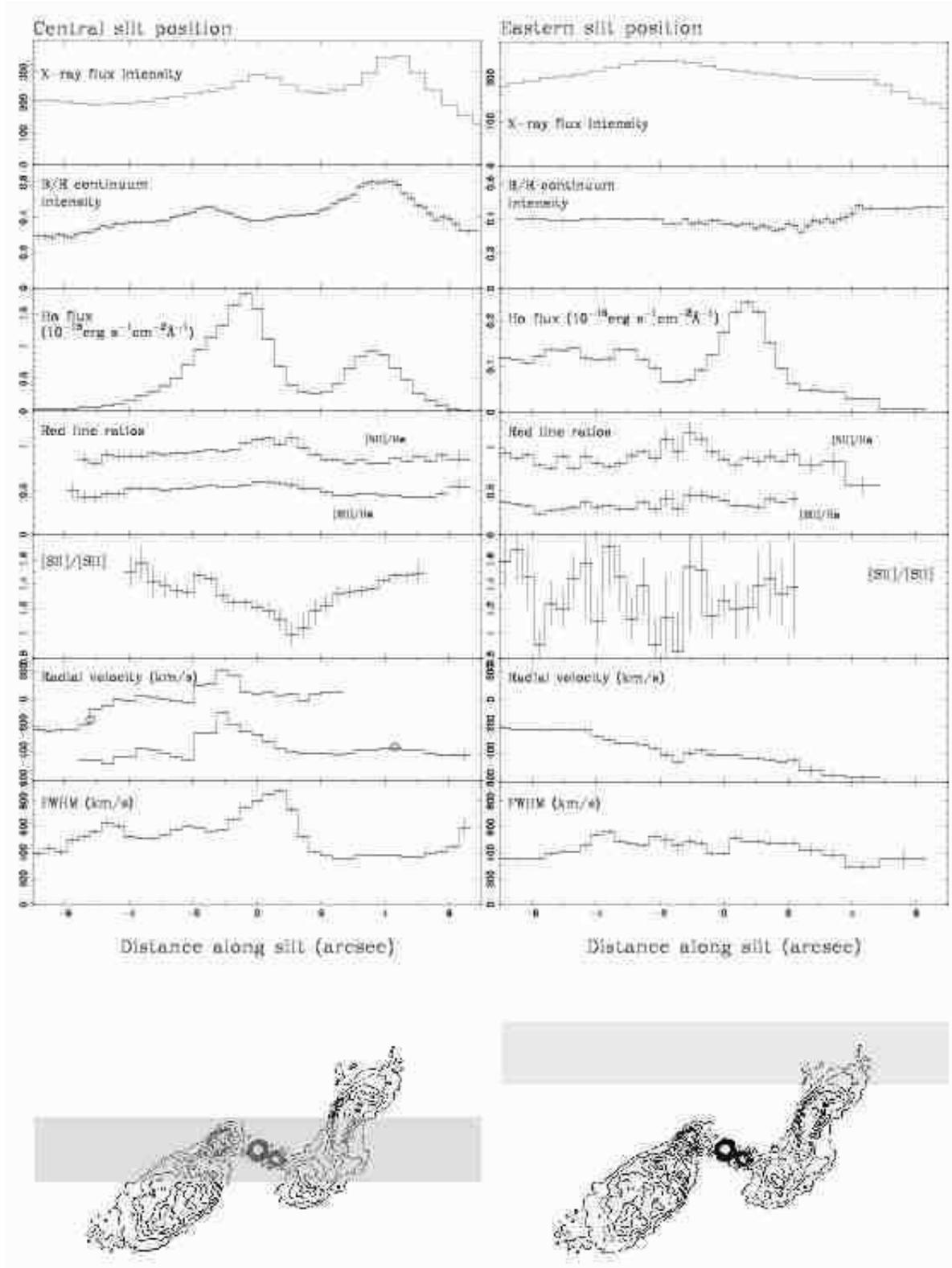,angle=0,width=0.90\textwidth}
\caption{ \label{fig:slitplotszoom}
The variations in the spectra along the central (left) and eastern
(right) slit positions, but now zoomed to the region local to, and relative to the radio source. 
Panels from top to bottom show: 
the X-ray flux intensity (in the 0.5-2keV band);
the ratio of the blue to the red continuum values (in the passbands
given in the caption to Fig~\ref{fig:cuts});
the intensity of the \ha\ emission line;
the intensity ratios of [NII]$\lambda6584$/\ha\ and [SII]$\lambda6717$/\ha;
the intensity ratio of [SII]$\lambda6717$/[SII]$\lambda6731$;
the radial velocity of the emission relative to the galaxy nucleus (the large open circles represent values estimated from the kinematic map of Hu \etal 1985); 
the FWHM of the \ha+[NII] lines (all when  modelled by only one velocity component);
and finally a scaled radio map (taken from Ge \& Owen 1993), with the slit size and position shown schematically by the grey box.   \newline 
In all figures, positive distances are to the north of the radio core, negative distances to the south. 
The $y$-axis scales differ 
between the left and right panels {\sl only} for the X-ray flux and the \ha\ 
flux intensities.}
 \end{figure}

The full-width at half-maximum (FWHM) of the emission lines is plotted
in Figs~\ref{fig:slitplots} and \ref{fig:slitplotszoom}; for most of
the eastern slit position this is fairly constant at around 430\kmps
across the nucleus and the inner part of the filament, dropping to
260\kmps beyond a distance of about 20~arcsec. The velocity width of
the filament emission seen through the central slit is highest at
large distances (300 to 400\kmps at 40 to 30~arcsec south), then declines
slightly to around 300\kmps over 28 to 10~arcsec south, thus appearing
fairly quiescent.  The FWHM increases sharply across the central
cluster galaxy however, showing three distinct peaks; two weak jumps
centred around 2.1 and 4.7~arcsec south, and a strong maximum centred
around 0.7~arcsec north, falling away sharply to be just less than
400\kmps beyond 1.5~arcsec north.

\onecolumn



\onecolumn
\twocolumn


\subsubsection{Stellar continuum fitting method} \label{sec:stars} We
model the stellar continuum observed across the galaxy and filament,
using the blue spectra from the central slit position only.  Fitting
is carried out within the IRAF SPECFIT package; individual fitting
components comprise the template non line-emitting central cluster
galaxy spectrum, along with five main-sequence stellar spectra (O5,
B5, A5, F5 and G5) to roughly characterize the excess blue continuum,
and K5V and K3III stars to help represent the old stellar population.
The stellar components were generated from the Kurucz (1979) tables
and the spectra were fit over the (rest-) wavelength range of
3300-5050\AA. The emission lines of [OII]$\lambda3727$,
[OIII]$\lambda\lambda4959,5007$, \hb\ and H$\gamma$ were also included
in the fit; as with the QDP fitting, we constrain [OIII]$+$\hb\ and
H$\gamma$ to have the same redshift and velocity width, and now also
permit them to be slightly skew in profile and have a small offset in
redshift from the stellar continuum. The procedure is iterative; we
start by fitting the template central cluster galaxy spectrum along
with the intensity of all emission lines, with the redshift, width and
skewness of the line freed after the first approximate fit is
obtained. We then add all the other stellar components in at an equal
weight (at around 1/10th of the level of the template galaxy) and
repeat the fitting iterations until there is no significant decrease
in $\chi^2$. No component is allowed to be negative, and any component
that is not required has its normalization automatically set to zero
in the fitting programme. 

To increase the signal of the spectra being modelled, we divide the
light from the central slit into eight distinct regions which comprise
(moving from south to north): \newline 

\noindent         {\sl a}) --36.5 to --28.3 arcsec (the outer filament);                        
\newline\noindent {\sl b}) --28.3 to --22.1 arcsec (the emission-line
peak just beyond the blue knots);  
\newline\noindent {\sl c}) --22.1 to --18.5 arcsec (the region of the
blue continuum knots);                           
\newline\noindent {\sl d}) --18.5 --7.7 arcsec (the inner part of the filament);                     
\newline\noindent {\sl e}) --7.7 to --2.7 arcsec (the southern flank of the central cluster galaxy);   
\newline\noindent {\sl f}) --2.7 to 0.2 arcsec (the southern blue lobe);                                        
\newline\noindent {\sl g}) 0.2 to 2 arcsec (the part of the galaxy lying inbetween the two peaks of the line emission );                
\newline\noindent {\sl h}) 2 to 6.3 arcsec (and the northern blue lobe).                                        

Our results are, however, subject to an uncertain correction for internal
reddening due to dust particles intrinsic to the galaxy.  Values of
the \ha/\hb\ Balmer decrement from a total spectrum across the central
galaxy suggest an average reddening of between E(B-V)=0.15 (within
$\pm5$\kpc) and E(B-V)=0.43 (within $\pm$13\kpc; Crawford et al 1999).
We do not apply such an {\sl averaged} value to our spectra, as the
HST (and other) images clearly show the dust to be structured into a
band of varying width (Fig~\ref{fig:zoomimages};  Pinkney \etal 1996) that does not
always intersect our slit positions. There is also little evidence
from these images for dust outside of the core of the dominant galaxy.
Whilst we could, in principle, estimate the intrinsic reddening from
\hb\ and H$\gamma$ Balmer decrement, both these emission lines are
affected by the Balmer absorption lines present in the blue stellar
continuum -- H$\gamma$ more so than \hb\ -- and the comparative
weakness in the H$\gamma$ makes the determination of the E(B-V) from
this ratio particularly uncertain.  We instead fit the continuum assuming {\sl no}
internal reddening to be present, and assume this represents a safe
{\sl minimum} contribution to the number and type of massive stars
present.

\subsubsection{Results of stellar continuum fitting}

\begin{figure}
\psfig{figure=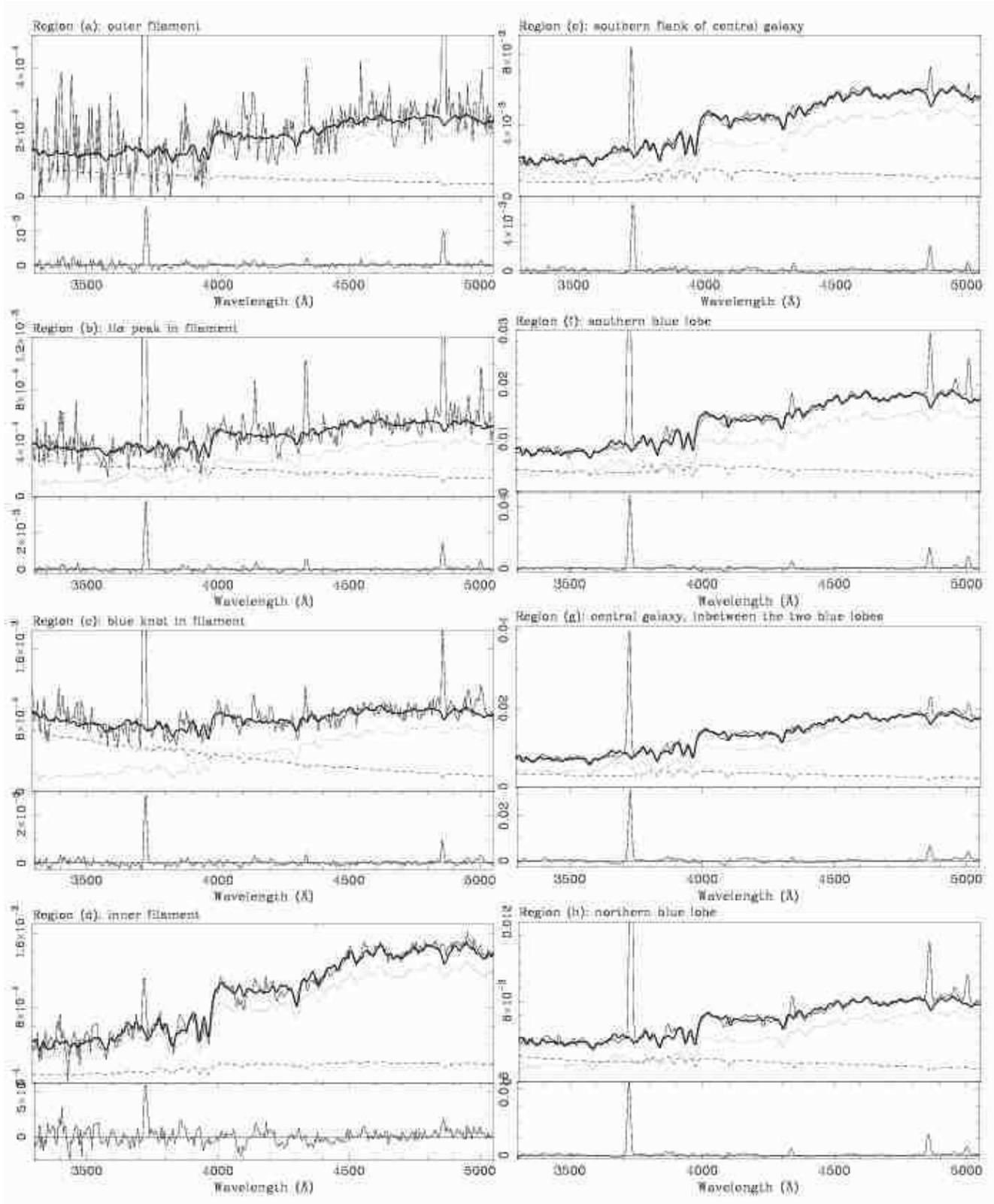,width=0.99\textwidth,angle=0}
\caption{\label{fig:stellarfits}
Stellar continuum fits to the spectra of the eight regions
(a) through (h) as detailed in the text, assuming {\sl no} intrinsic
reddening to be present. The best fit presented in
Table~\ref{tab:stars} is shown as a thick solid line superposed on the
data (thin solid line) in the upper panel; the residual between this
model and the data is shown in the panel below each plot. The
late-star component (K5V$+$KIII$+$template central cluster galaxy) is
shown as a dotted line, and the young star component
(O5V$+$B5V$+$A5V$+$F5V$+$G5V) is shown as the dashed line, both in the
upper panel.  The
$y$-axis gives the intensity in units of 10$^{-15}$\ergpspcmsqpA.
Note that the upper panels are scaled in the $y$ direction to show the
continuum features, whilst the lower panels are scaled so as to show
the relative strengths of the emission lines. 
}
\end{figure}

The resulting decomposition of our best model fits into early- and
late-type stellar components are given in Table~\ref{tab:stars}. The
components are all represented as percentages of the total light at
(rest-frame) 4400\AA, and the total flux in the spectrum at that point
is given in the final column.  The best fits to the spectra are also shown in
Fig~\ref{fig:stellarfits}.  

Despite being up to 37 arcsec away from the core of the galaxy, all
the spectral regions require a component of older stars, as modelled
by the template galaxy and the K5V and KIII stars.  Such a late-type
component at large radii should not be surprising, however, given the extensive
cD envelope known to surround this central dominant galaxy (Johnstone 
et al 1991).  This is reinforced by the fact that the spatial profile
of the 4500-4800\AA\ continuum shown in Fig~\ref{fig:cuts} is non-zero
out to beyond 30 arcsec south.  The decrease of the flux in the
4500-4800\AA\ band with radius matches well the fall in the flux of
the old stellar component in the models with radius. As can be seen in
Fig~\ref{fig:stellarfits}, the old stellar component in all the models
is very constant at accounting for between 70 and 85 per cent of the
flux in this band.  Thus the \lq old' stellar component in all of the
model fits appears to be due to the large extent of the underlying
galaxy. 


The excess blue continuum clump in the filament (region {\sl c}) is
almost entirely comprised of O stars, suggesting an episode of either
very recent or ongoing star formation. Interestingly, the two regions
in the filament further out from the nucleus (regions {\sl a} and {\sl
b}, coincident with the \ha\ peak and beyond) show a spectrum dominated
by O star emission, although with a smattering of lower-mass stars
also present. The inner half of the filament shows little in the way
of young stars, and the northern and southern blue lobes have an older
population of stars than seen in the star formation region in the
filament. We note, however, that it is precisely regions {\sl f}, {\sl
g} and {\sl h} that would be most affected by dust extinction, so we
expect that de-reddening the spectra would show the spectrum in these
regions to be likewise dominated by younger stars. Dust extinction is
unlikely to affect the star formation in the filament. 

\subsubsection{Results: the blue line intensities and ratios}
\label{sec:revisitlines}
\begin{figure}
\psfig{figure=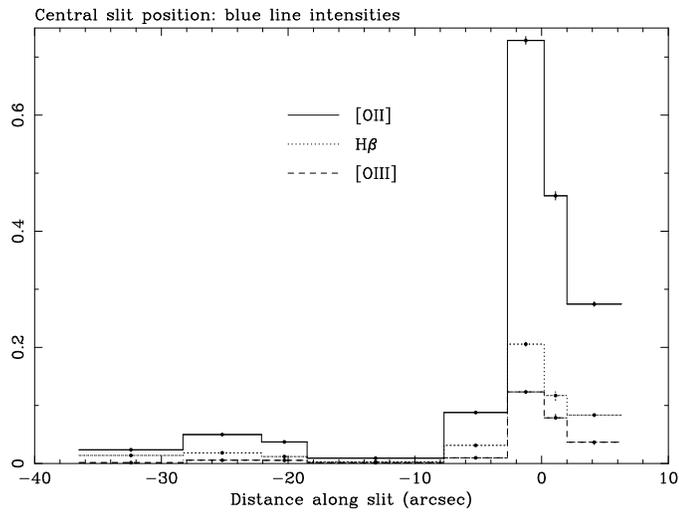,width=0.5\textwidth,angle=270}
\caption{ \label{fig:bluelinesfit}
The intensities of the main blue emission lines of [OII]$\lambda$3727
(solid line),
\hb\ (dotted line) and [OIII]$\lambda$5007 (dashed line) along the central slit position.}
\end{figure}

\begin{figure}
\psfig{figure=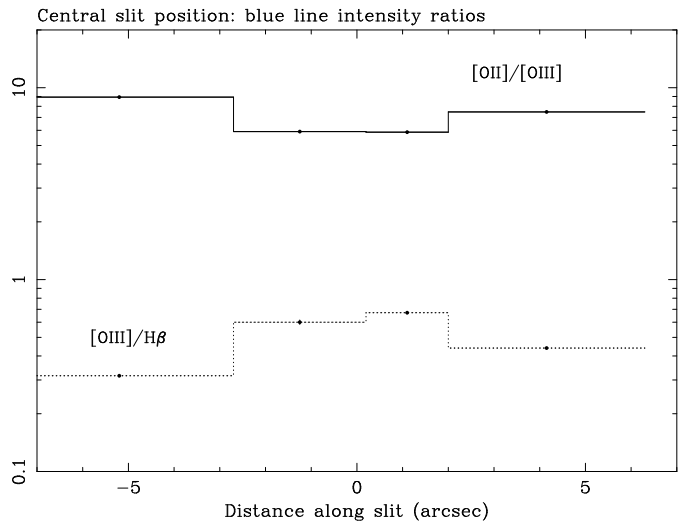,width=0.5\textwidth,angle=270}
\caption{ \label{fig:bluerats}
The emission line intensity ratios of [OII]/[OIII] (solid line) and [OIII]/\hb\
(dotted line) across the dominant galaxy and contained radio source,
taken from the central slit
position. }
\end{figure}

We briefly revisit the properties of the blue emission lines, now
corrected for any underlying stellar Balmer absorption. The variation
in the line intensity along the central slit
(Fig~\ref{fig:bluelinesfit}), as expected, mimics the distribution
seen in the \ha\ intensity (Fig~\ref{fig:slitplots}). The blue line
intensity ratios, like the red line ratios, show a sharp change in
value across the region of the slit encompassing the radio source
(Fig~\ref{fig:bluerats}; note that the absolute value of [OII]/[OIII]
is still probably inaccurate though, as we have not made any
correction for intrinsic reddening). 

\onecolumn

\twocolumn 

\onecolumn
\begin{table}
\caption{Stellar synthesis results}
\label{tab:stars}
\begin{tabular}{lcrrrrrrrrl}
\hline
Region &template & O5v  & B5-7v & A5v & F5v  & G5v  & K5v & K3iii & Flux at 4400\AA & Description \\
       &galaxy (\%) & (\%) & (\%) & (\%) & (\%) & (\%) & (\%) & (\%) & ($10^{-18}$\ergpspcmsqpA) & \\
\hline
{\sl a }& 69 & 15 &  --  & -- &   4  &  6   &  4   & 2    &  0.19 & outer filament\\
{\sl b }& 56 & 18 & $<$1 &  2 &  11  &  4   &  3   & 5    &  0.49 & \ha\ peak in filament \\
{\sl c }& 64 & 32 & --   &  3 & --   &  1   & $<$1 & $<$1 &  0.80 & blue knots in filament\\
{\sl d }& 69 & -- & $<$1 &  5 & $<$1 & 12   &  13  & $<$1 &  1.11 & inner filament \\
{\sl e }& 69 &  1 &  2   & 13 &   9  & $<$1 & 2    &  3   &  4.71 & southern flank of central galaxy\\
{\sl f }& 73 &  4 &  7   &  5 &  11  &  --  & --   & --   & 14.42 & southern blue lobe\\
{\sl g }& 79 &  3 & 12   &  2 &   4  &  --  & --   & --   & 14.14 & galaxy inbetween blue lobes\\
{\sl h }& 73 &  9 &  6   &  3 &   9  &  --  & $<$1 & --   &  4.86 & northern blue lobe\\
\hline
\end{tabular}
\\
The entries in the table show the percentage contribution by each component to the
total light at 4400\AA, except for the last column which gives the observed
flux of the spectrum at that same point. 
\end{table}

\twocolumn

\section{Discussion}

\subsection{The immediate environment of the radio jet} 
\subsubsection{The northern radio lobe} As it emerges from the active
nucleus, the northern jet of radio plasma appears to displace
the intracluster medium.  The spatial distribution of both the X-ray
gas and the warm ionized gas giving rise to the optical emission lines
both show a clear dip in their intensity between 1-2.5 arcsec north in
both the images (Fig~\ref{fig:zoomimages}) and in the spectra from
along the central slit position (Fig~\ref{fig:slitplotszoom}). This
may well be in part due to the location of the dust lane, which seems
to have also been similarly displaced in this direction, along with
the molecular gas (Fig~\ref{fig:zoomimages}). 

After this displacement -- or maybe because of this displacement, if
it results in compression of the intracluster medium -- the northern
radio jet then clearly encounters a dense obstruction at around 2.5
arcsec to the north-west of the active nucleus, causing its sharp
deflection to the north-east. Evidence for a direct interaction
between the radio jet and a dense cloud is found from the optical
spectra over the region of 0--2 arcsec north of the active nucleus in
the central slit position. Within this region -- coincident with a
bright radio knot -- there is a noticeable increase in the FWHM of the
emission lines, and a very sharp drop in the [SII]/[SII] line ratio
from the low-density values implying a hundred-fold increase in the
electron density.  Other line ratios, such as the [NII]/\ha,
[SII]/\ha, [OII]/[OIII] and [OIII]\hb\ also change to a state of
higher ionization in the same region.  The spectra thus imply that
here that the radio plasma pushes out against the surrounding gas,
increasing its ionization state, turbulence and density. 

Beyond this point of collision, bright peaks are apparent in the X-ray
flux and optical and CO line emission all lying to the outside of the
deflected radio lobe.  The optical line emission from this peak is no
longer turbulent (FWHM of $\sim400\kmps$) but it has a strong
blueshift of around $-400$\kmps, implying that this gas has had a
momentum boost from a collision with the radio jet. The CO line
emission is also observed to have a similar strong blueshift (of the
order of $-300$\kmps in this region; Salom\'e \& Combes 2004).  The
Eastern slit position crosses the radio source only at the outer
reaches of the northern lobe, and the peak \ha\ emission here is
complementary to the dip in the line emission seen in the central
slit. The velocity of all this emission line gas to the east is
substantially blueshifted (up to almost --600 \kmps relative to the
nucleus) suggesting that this gas has also been
displaced by the radio source, and forms the the eastern end of the
blueshifted component seen in the central slit position. 

Along our central slit position, the X-ray and \ha\ peaks clearly lie
at the edge of the radio lobe (Fig~\ref{fig:slitplotszoom}), along
with the peak in the blue starlight. The peaks in these three
components are not
exactly spatially coincident with the CO peak, which is centred around
3.3 arcsec distant from the active nucleus, whereas the \ha\ peak lies
at 3.6$\pm$0.2 arcsec and the X-ray peak is furthest out at
4.25$\pm$0.5 arcsec.  The blue starlight lies between \ha\ and X-ray,
centred at around 3.8 arcsec. The CO peak is also slightly
more west (by about 13 degrees) than the other two peaks
(Figs~\ref{fig:zoomimages} and \ref{fig:slitplotszoom}), thus lying
closer to the point of impact. 

Given the general Z-shaped morphology of the source, it is possible
that the radio jets are precessing or oscillating, and thus that the
point of impact with the surrounding intracluster medium has been
changing with time.  The CO peak could then mark the most recent point
of impact, where the gas has been compressed but star formation has
yet to be initiated. The region of brighter X-ray and \ha\ emission
might then be associated with an earlier point of contact. Compression
of the gas has triggered the observed massive recent star formation,
either as a single young starburst or as a continuous and longer-lived
period of star formation (Smith et al 1997).  The actual interaction
may have also caused some mixing of the hot intracluster medium and
the warm molecular gas, resulting in soft X-ray and UV ionization that
in turn is capable of ionizing the cold clouds to produce the optical
line emission in these regions (in addition to any stellar ionization).  

\subsubsection{The southern radio lobe}
The other radio lobe also has a clear kinematic interaction with the
warm ionized gas.  Emission from the filament dominates south of about
$-5$ arcsec of the active nucleus, but then a second component of gas
(as evinced by its different, and blueshifted velocity) becomes
apparent just at the outer edge of the southern radio lobe.  It is at
this point, that the first noticeable increase in linewidth occurs.
The (redder) emission from the \lq filament' component of gas dominates 
in to the core of the galaxy, but the second blueshifted component 
continues right across the galaxy, becoming very prominent to the
north, eventually forming the north-west peak. The blueshifted
line emission is thus entirely associated with the location and
properties of the radio source. 

Both velocity components show a marked jump in redshift (by 200\kmps
to the red) at $-2$ arcsec, just beyond the location of the sharp
deflection in the southern radio lobe at --1~arcsec
(Fig~\ref{fig:slitplotszoom}).  The pronounced increase in the FWHM
(and decrease in the [SII]/[SII] ratio) starts within this region,
reaching a maximum (minimum) to the north of the nucleus, before the
radio lobe deflection point on that side. The distribution of the \ha\
emission seen in our central slit has a very skewed distribution to
the south of the nucleus, peaking at --3.5$\pm$0.8 arcsec. Thus,
unlike with the northern lobe, it does not seem to have been
completely pushed out to lie only along the outside of the radio,
although a tail in this peak spans the whole southern radio lobe. The
multiwaveband images suggest that there are is also components of
optical line emission, CO line emission, and UV light that appear to
be extended a short way {\sl along} the direction of the southern
radio lobe (Figs~\ref{fig:zoomimages} and ~\ref{fig:images}). In
contrast, there is X-ray emission across the nucleus and beyond the
point of deflection of the southern radio jet, but there is, if
anything, a deficit of X-rays associated with the southern radio lobe
location.  Along our central slit position, the star formation peaks
at --1.5 arcsec, again just at the outside of the southern radio lobe. 

Thus although the kinematics of the gas suggest that there is some
input of turbulence and energy into the warm gas from the southern
radio jet, which has in turn triggered star formation, the effects on
the distribution and ionization state of the gas are far less
pronounced than for the northern jet-cloud interaction. 

\subsection{The filament } 

The radio source has a marked influence on the surrounding environment
only within about $\pm6$ arcsec of the active nucleus.  Beyond that,
the soft X-ray filament stretches to almost 50~arcsec south of the
dominant galaxy, and could show where the X-ray
intracluster medium is cooling around the gravitational wake of the
central cluster galaxy as it oscillates through the core of the
cluster (Fabian et al 2001). This gives us a direct view of cooling
within the intra-cluster medium well-removed from the influence of the
dominant galaxy and its contained radio source. 

In such a situation, the densest regions start to cool most rapidly,
becoming the first to collapse and form massive stars (in the absence
of any other heating mechanism).  The chain of resolved 
continuum knots centred
around --21~arcsec south presumably are stellar clusters resulting
 from the early collapse of
the densest part of the filament.  As the light from this region is
comprised almost entirely of O stars (Table~\ref{tab:stars}; Mittaz et
al 2001), the star formation occurred within the last few tens
of millions of years. The strongest clumps of X-ray emission within
the filament (centred at $-$9.8, --22.5 and --27.5~arcsec south of the
nucleus) thus trace the less dense regions along the filament that
have been cooling more slowly. The strongest \ha\ emission peak along
the filament coincides with the X-ray peak at around --23~arcsec
south.  [The field of view of the CO line observations of the central
galaxy does not include the filament, so we cannot comment on the
location of any molecular gas components along the filament.]

Our simple spectral synthesis shows that the total 
light (at 4400\AA) from
the UV knots within the slit is due to over 100
main-sequence O5 stars.  Mittaz et al (2001) deduce over ten times
more O stars from the UV flux of this region, but they assume an
intrinsic reddening appropriate to the central cluster galaxy
($E(B-V)\sim0.14$), which is unlikely to be appropriate out in the
filament. The inferred \ha\ luminosity produced by recombination in
gas illuminated by 100 O5 stars is $6\times10^{38}$\ergps (scaling
from Allen 1995) a factor of over 100 below the observed \ha\
luminosity from the same region of the slit. This discrepancy is
exacerbated by the significant spatial offset of $\sim$2~arcsec between
this region of star formation and the brightest peak of the \ha\ line
emission. Thus the stars not only fail to provide enough ionization,
but they are clearly not in the right location to provide the
ionization for the optical line emission. 

The bulk of the filament (between $-$40 to --10 arcsec south) is
blueshifted only by around $-50\pm50$\kmps with respect to the central
cluster galaxy. Its fairly constant velocity width of 
$\sim300\pm100$\kmps implies an r.m.s. velocity of the gas within the
filament of around $130\pm50$\kmps of its mean radial velocity. Thus
despite the filament originating from the motion of the dominant
galaxy, the gas motions within the filament are not large and
turbulent. It is hence unlikely that mixing layers are relevant as a
source of ionization. The fact that the soft X-ray luminosity from the
filament is only an order of magnitude more luminous than the
\ha+[NII] emission (Fabian et al 2001) compounds the problem.  As seen
in many of these cooling flow nebulae it is clear that some -- and as
yet undetermined -- source of heat is required to power the optical
line emission. Perhaps there is a small amount of energy conducted
inwards from the surroundings, or by energetic electrons particles
from the hot gas (depending on the magnetic configuration of the
filament).  

The UV knots are located around 24~\kpc\ south of the current position
of the galaxy.  If the galaxy is moving through the cluster at a
relative rate of $+374$\kmps (Oergerle \& Hill 1994) it would take
approximately 66 million years to travel between the UV knots to its
present location (or approximately 160 million years if travelling at
$+150$\kmps). This timescale fits well with the expected lifetime of
the star clusters as their light is currently dominated by massive young O
stars. A further consequence is that the optical line emission will
itself have been around for at least this long, as it will have taken
the galaxy around 120 million years to travel its full length (or
$\sim300$ million years for the slower speed).  The length of the
filament also shows that the warm emission-line gas is not mixing
rapidly in with its surroundings; it is therefore a
long-lived feature of the core of this cluster of galaxies. 

The accumulated data are therefore consistent with a cooling wake
interpretation, provided that gas really is cooling. The \c\ data
allow for about $100\Msunpyr$ cooling from the inner intracluster
medium, but we see less than $1\Msunpyr$ of star formation in the giant
filament. The situation in A\,1795 resembles that in many cluster cores
where the radiative cooling time of the gas is short. The gas
temperature ranges down from the outer virial temperature by a factor
of about 3 (Peterson et al 2001, 2003) in a manner consistent with the
gas cooling, but the gas is not obviously accumulating at that lower
temperature nor seen at still lower temperatures. Unless there is an
unseen sink of cooled gas, most of the gas must be heated and
prevented from collapsing. The exact heating mechanism is not yet
identified, although it is likely that the central radio source is
involved (e.g. Fabian 2005 and references therein).

An alternative interpretation is that the giant filament consists of
cold gas from the central galaxy which has been dragged out. A set of
filaments over 40~kpc long is seen going to the North of NGC\,1275 in
the Perseus cluster (Lynds 1970; Conselice et al 2001). These have
mostly probably been dragged out by a buoyant bubble made by the
central radio source (Fabian et al 2003). A \lq horseshoe' shaped
filament lying just inside the well-known ghost bubble in the Perseus
cluster supports this idea. The small velocity range found here in the
giant filament in A\,1795 indicates that gas flows must be fairly smooth
and laminar and not highly turbulent.

\section{Conclusions}

The changes in line intensity ratios, radial velocity and velocity
widths of the optical emission-line gas provide further evidence
supporting the generally-accepted scenario of an interaction between
the radio jet and its surrounding intracluster medium, inducing recent
massive star formation within the central cluster galaxy.  We resolve
the light from young star clusters comprised in total of at least 100
O stars, halfway along the filament. At an age of only a few million
years, and at a distance of 21\kpc\ from the central galaxy, these
stars are unlikely to have been dragged out of the galaxy.  Thus it
seems that even well away from the central galaxy and any influence of
its radio source, straightforward cooling of the intracluster medium
within the filament has resulted in the recent production of 
modest-sized star clusters. These stars cannot power the \ha\
line-emission which requires an (as yet) undetermined heat source.
Despite the input of this heating, the filament of A\,1795 shows that
some of the intracluster medium is able to cool all the way from the
hot X-rays to condense to form massive stars.  The presence of these
star clusters within a cooling filament lends weight to the possibility
of a similar origin for the young and luminous star clusters found
around the central galaxy NGC1275 in the Perseus cluster of galaxies
(Holtzmann et al 1992; Richer et al 1993).

\section*{Acknowledgments}

CSC and ACF thank the Royal Society for financial support.  The
William Herschel Telescope is operated on the island of La Palma by
the Isaac Newton Group in the Spanish Observatorio del Roque de los
Muchachos of the Instituto de Astrofisica de Canarias

{}

\label{lastpage}

\end{document}